\documentclass{aa}
\bibpunct{(}{)}{;}{a}{}{,}
\usepackage{natbib}
\pdfoutput=1
\usepackage{multirow}
\usepackage{amssymb}
\usepackage{amsmath}

\usepackage{float}
\usepackage{subfig}
\usepackage{graphicx}
\usepackage{txfonts}
\begin{document} 
\titlerunning{Exploring the multiphase medium in MKW 08}
\title{Exploring the multiphase medium in MKW 08: from the central active galaxy up to cluster scales}
  \authorrunning{A. T\"{u}mer et al.} 
   \author{A. T\"{u}mer \inst{1,2} \and F. Tombesi \inst{2,3,4,5} \and H. Bourdin \inst{2} \and E. N. Ercan \inst{1} \and M. Gaspari \inst{6,7} \and R. Serafinelli \inst{8}}
     
   \institute{Department of Physics, Bo\u{g}azi\c{c}i University, Bebek, 34342 Istanbul, Turkey\\
              \email{aysegul.tumer@boun.edu.tr}
         \and
             Department of Physics, University of Rome "Tor Vergata", Via della Ricerca Scientifica 1, 00133 Rome, Italy
                      \and
             Department of Astronomy, University of Maryland, College Park, MD 20742, USA
                      \and
             NASA/Goddard Space Flight Center, Code 662, Greenbelt, MD 20771, USA
                      \and
             INAF Osservatorio Astronomico di Roma, Via Frascati 33, 00078 Monteporzio Catone, Italy
             \and
             Department of Astrophysical Sciences, Princeton University, 4 Ivy Lane, Princeton, NJ 08544-1001, USA
             \and
             Lyman Spitzer Jr. Fellow
             \and
              INAF Osservatorio Astronomico di Brera, Via Brera 28, 20121 Milan, Italy}
             %\thanks{The university of heaven temporarily does not accept e-mails}

\date{Received April 10, 2019; accepted August 16, 2019}

% \abstract{}{}{}{}{} 
% 5 {} token are mandatory
 
  \abstract
 {The study of the brightest cluster galaxy (BCG) coronae embedded in noncool core (NCC) galaxy clusters is crucial to understand the BCG's role in galaxy cluster evolution as well as the activation of the self-regulated cooling and heating mechanism in the central regions of galaxy clusters.}
   {We explore the X-ray properties of the intracluster medium (ICM) of the NCC galaxy cluster MKW 08 and the BCG corona, along with their interface region. With recent and deep archival \textit{Chandra} observations, we study the BCG corona in detail, and with archival \textit{XMM-Newton} observations, we investigate the implications of the central active galactic nuclei (AGN) on the BCG.}
   {We carry out imaging and spectral analyses of MKW 08 with archival \textit{XMM-Newton} and \textit{Chandra} X-ray observations.}
{Our spectral analysis suggests the presence of a central AGN by a power-law with a photon index of $\Gamma$ $\simeq$ 1.8 at the core of its BCG. Although the ICM does not exhibit a cluster scale cool core, the BCG manifests itself as a mini cool core characterized by a cooling time as short as 64 Myr at \textit{r} = 3 kpc centered at the galaxy. The isothermality of the BCG corona seems to favor mechanical feedback from the central AGN as the major source of gas heating. The gas pressure profile of this mini cool core suggests that the BCG coronal gas reaches pressure equilibrium with the hotter and less dense ICM inside an interface of nearly constant pressure, delimited by radii 4 $\leqslant$ \textit{r} $\leqslant$ 10 kpc at the galactic center. As revealed by the presence of a metal enriched tail (\textit{Z} $\simeq$ 0.5 - 0.9 \textit{Z$_{\odot}$})  extending up to 40 kpc, the BCG corona seems to be experiencing ram-pressure stripping by the surrounding ICM and/or interacting with a nearby galaxy, IC 1042.}
  {}


\maketitle
%
%-------------------------------------------------------------------

\section{Introduction}

Clusters of galaxies are the largest gravitationally bound structures in the universe. Elements produced inside a galaxy cluster can rarely escape its deep gravitational potential well, therefore these constituents make clusters great probes for understanding the evolutionary history of their galaxies, the feedback from supermassive black holes (SMBH), chemical enrichment, and stellar evolution. 

The intracluster medium (ICM) is an optically thin hot plasma ($\sim$10$^{7}$-10$^{8}$ K) that fills in between the galaxies inside clusters. It accounts for the $\sim$12\% of the baryonic matter inside galaxy clusters and its emission prevails in the X-ray band of the electromagnetic spectrum \citep[see, e.g.,][]{Sarazin86}. This plasma is almost spherically symmetric and close to hydrostatic equilibrium in relaxed clusters. Brightest cluster galaxies (BCGs) are the most massive galaxies residing at the center of the cluster gravitational potential well, and they most likely affect the surrounding ICM \citep[see, e.g.,][]{Sun09}. In return, the evolution of BCGs is likely linked to their host clusters \citep[see, e.g.,][]{Lin07,McDonald18}. Compared to other galaxies in a cluster with similar stellar and black hole mass, BCGs hold a higher fraction of radio-loud active galactic nuclei (AGN) \citep{Best07} that are extremely bright due to the accretion of matter around a central SMBH \citep[see, e.g.,][]{Sadowski17}.

X-ray emitting processes in the ICM are mainly in the form of thermal bremsstrahlung and of line emission. Due to this thermal emission, the ICM cools down proportionally to the square of the gas density. Therefore, the cooling is more prominent in higher density regions, mainly in the central regions of galaxy clusters. Cooling flows could develop inside these regions, characterized by a cooling time much shorter than the Hubble time. However, it is recognized that the impact of such cooling flows is much smaller than anticipated \citep{Molendi01}. Thus, some heating mechanisms must be balancing any excessive cooling of the gas residing in clusters cores.

Recent studies suggest that AGN feedback may influence the thermodynamics of cluster atmospheres at  $\textit{r}$ < $0.2$ R$_{500}$ \footnote{R$_{\Delta}$ corresponds to the radius of a point on a sphere whose density equals $\Delta$ times the critical density of the background universe, $\rho$$_{c}$(z), at the cluster redshift where $\Delta$ is the density contrast.} and contributes to the regulation of ICM cooling via the injection of mechanical energy \citep[see, e.g.,][]{Gaspari14}. In exchange, without the matter feeding them, the growth of SMBHs would not be possible \citep[see, e.g.,][]{Gaspari17}. This establishes a feeding and feedback loop between the central AGN and the ICM. Although the exact mechanism of the cycle of heating and cooling is still debated, the interaction of central AGNs and the ICM in the form of large-scale shocks and ICM bubbles is confirmed by numerous studies \citep{Birzan04,Fabian03,McNamara05,Nulsen05}. Furthermore, there is  a strong relation between the SMBH inside the galaxies and the constituents of their interstellar medium (ISM) by means of multiscale and multiphase outflows \citep[see, e.g.,][]{Tombesi15,Cicone18,Serafinelli19}. Yet, how these mechanisms affect the evolution of clusters remains to be better understood.

Using \textit{ROSAT} All-Sky Survey, \citet{Reiprich02} use a sample of 64 galaxy clusters with a minimum flux limit of 2 $\times$  10$^{-11}$ erg s$^{-1}$ cm$^{-2}$ in the 0.1-2.4 keV energy band that is known as HIghest X-ray FLUx Galaxy Cluster Sample (HIFLUGCS). This sample constitutes clusters that are categorized by their central cooling time at 0.4\% R$_{500}$, as strong-cool core (SCC,  t$_{cool}$ < 1.0 Gyr), weak-cool core (WCC, 1.0 < t$_{cool}$ < 7.7 Gyr), and noncool core (NCC,  t$_{cool}$ > 7.7 Gyr) clusters \citep{Hudson10}. The HIFLUGCS clusters are at 44\% of SCC type, where 28\% are of WCC, and the remaining 28\% are identified as NCC. All of the cool core (CC) types of clusters host an AGN in their BCGs, whereas only 45\% of NCC clusters in this sample show the presence of a central AGN \citep{Mittal09}. CC clusters show a radial temperature distribution with a low central temperature that rises gradually away from the center, then shows a gradual decrease toward the outskirts. NCC clusters, on the other hand, do not exhibit any prominent decrease in their central temperature \citep[see, e.g.,][]{Molendi01}. 

However, the distinction between CC and NCC types of clusters seems to be more complex. The launch of \textit{Chandra} led to the discovery of small scale cool cores ($\sim$ 3 kpc) of two dominant galaxies inside an NCC cluster Coma, where the interstellar gas is found to be confined by and in pressure equilibrium with the hotter ICM \citep{Vik01}. In addition, the interstellar gas temperatures of these coronae are reported to be in the range of \textit{kT} = 1-2 keV by the authors. These findings paved the way for a systematic search for similar objects in NCC clusters, which behave as mini versions of cluster cool cores.

Following the discovery of numerous other X-ray thermal coronae by \citet{Sun07}, instead of CC - NCC classification, \citet{Sun09} points to a dichotomy of a large cool core (LCC) class with their gas content of ICM origin and a characteristic scale of \textit{r}$_{4Gyr}$ > 30 kpc, and a corona class where the gas is of ISM origin with a typical radius of \textit{r} $\leqslant$ 4 kpc. The distributions of these classes are separated by their luminosities at the value \textit{L$_{X}$} $\sim$ 4  $\times$ 10$^{41}$  erg s$^{-1}$.  Various clusters that were previously identified as NCC, are proved to have coronae with luminous radio-loud AGNs \citep{Sun09}. \citet{Sun07} discuss distinctive features of these embedded X-ray thermal coronae in 25 nearby hot (3 keV < \textit{kT}) clusters that behave as mini cool cores. These coronae can be identified with a prominent iron L-shell bump, and they show extended soft X-ray emission. The temperature values of this sample of embedded coronae vary in the range of 0.3 < \textit{kT} <1.7 keV and the abundance of the coronal gas is found to be \textit{Z} $\sim$ 0.8 \textit{Z$_{\odot}$}. The gas origin and the evolution of these coronae suggest that these coronae are formed prior to the evolution of their host galaxies within their host clusters. They seemed to have survived ICM stripping since once they are destroyed and their interstellar space is filled with the hot ICM, they would not be able to reform a cool core. In addition, they suggest that these cool coronae may be the kernels for large cluster scale cool cores. In addition, ISM temperatures of isolated galaxies also reside between 0.3 < \textit{kT} <1.2 keV \citep{Xu02,Werner09,Kim12}.

The poor galaxy cluster MKW 08 is part of the HIFLUGCS with redshift \textit{z} = 0.027 \citep{Lin04}, which is classified as a NCC cluster due to its central cooling time $t_{cool}$ = 10.87 Gyr \citep{Hudson10} at \textit{r} = 0.4\% $R_{500}$, and it has a central radio source \citep{Bharadwaj14}. Its core metallicity is found to be \textit{Z} $\sim$ 0.5 \textit{Z$_{\odot}$} with an overall temperature value of \textit{kT} $\sim$ 2.5 keV \citep{Elkholy15}. The BCG of MKW 08 is NGC 5718, as reported by \citet{Lin04}. The cluster is gravitationally bound with the group MKW 07 and possibly heading for a merger in several Gyrs \citep{Beers95}. In HIFLUGCS Cosmology Study (HICOSMO) by \citet{Schellenberger17}, the temperature profile of MKW 08 shows a typical NCC trend with gradual decrease from the central region to $R_{500}$ by extrapolation. This trend is also seen in the study by \citet{Bharadwaj14} where they use a region of $\sim$ 2$\arcmin$ for their innermost bin, which may have hidden the smaller scale structures and resulted in the overestimation the central temperature ($\sim$ 3.7 keV). 

MKW 08 is reported to have an X-ray corona inside its BCG by \citet{Sun07} using a single \textit{Chandra} observation where the lack of data quality (both in shallow exposure time and the fact that BCG was located at a chip gap) limited the authors to thoroughly study this region. In this work, we consider two more recent deep \textit{Chandra} archival observations to resolve thermodynamical profiles of the \textit{X-ray corona} of NGC 5718 on kpc scales. Thanks to their arcsecond angular resolution, the \textit{Chandra} ACIS-I pointings allowed us in particular to locate the region of interaction of the BCG with the nearby interacting galaxy IC 1042 \citep{Hudson10}. We complemented our analysis using the \textit{XMM-Newton} EPIC observations, that are characterized by twice the larger field of view than the \textit{Chandra} ACIS-I cameras, and provided us with a higher effective area at high energy. \textit{XMM-Newton} EPIC pointings especially helped us to investigate thermodynamical properties of the ICM at larger scales, and to best characterize the spectral properties of the central AGN. We focus on the BCG of MKW 08, NGC 5718, specifically since this object is a rare example of a deep \textit{Chandra} observation of a BCG corona embedded in the ICM of a noncool core cluster, which allows the characterization of the physical phenomena related to the \textit{X-ray corona} and the ICM cooling and feedback.

This work is structured as the following. In Section~\ref{sec:Obs}, we describe the \textit{XMM-Newton} and \textit{Chandra} archival observations of MKW 08, data reduction methods as well as the background modeling. In Section~\ref{sec:DataAnalysis}, we obtain the proper cluster emission to construct temperature and surface brightness maps, then we present the radial surface brightness profiles fit with a double $\beta$-model for projection, and estimate the radial 3-dimensional temperature and pressure profiles of the cluster. Then, we focus on the spectral analysis of the central BCG, NGC 5718, in Section~\ref{sec:NGC5718} using both \textit{XMM-Newton} and \textit{Chandra} observations where we also present the luminosity estimates and radial cooling time profile. The spectroscopic properties of a $\sim$ 40 kpc BCG tail is investigated in Section~\ref{sec:Tail}. In the last two sections, we discuss and summarize our findings. 

Throughout this paper, we assume the $\Lambda$CDM cosmology with \textit{H$_{0}$} = 70 km s$^{-1}$ Mpc$^{-1}$, $\Omega$$_{0}$ = 0.3, 
$\Omega$$_{\Lambda}$ = 0.7. According to these assumptions, a projected intracluster distance of 100 kpc corresponds to an angular separation 
of  $\sim$ 180 $\arcsec$. Unless otherwise stated, confidence levels on the parameter estimates, confidence envelopes and error bars correspond to 
68\%. The redshift value is fixed at \textit{z} = 0.027 \citep{Lin04} and we adopt the mean atomic hydrogen column density value of \textit{N$_H$} 
= 2.45 $\times$ 10$^{20}$ cm$^{-2}$ retrieved from the \textit{N$_H$} calculator tool \footnote{This calculator is accessible via 
www.swift.ac.uk/analysis/nhtot/.} based on the methods of \citet{Willingale13} and the data of The Leiden/Argentine/Bon (LAB) Survey of Galactic 
HI \citep{LAB}. Within the scope of this work, we study the X-ray corona of NGC 5718 and we refer it to as "the corona" throughout the text.

\section{Observations and data reduction}
\label{sec:Obs}

\subsection{Observations}
\begin{table*}
\caption{Effective exposure time of each \textit{XMM-Newton} EPIC and \textit{Chandra} ACIS-I observation.}
\label{tab:obslog}   
\centering
\begin{tabular}{c c c c c}
\hline\hline
\textit{XMM-Newton} & Equatorial coordinates & MOS1 effective  &MOS2 effective & PN effective  \\  
Observation ID &(J2000) &  exposure time (ks) & exposure time (ks) & exposure time (ks) \\
\hline
        0300210701 & 14 40 38.1 +03 28 18.1 &23.1 & 23.1 & 16.5\\            
\hline
\textit{Chandra} & Equatorial coordinates & \multicolumn{3}{c}{ ACIS-I effective}  \\  
Observation ID&(J2000) & \multicolumn{3}{c}{exposure time (ks) } \\
\hline
        4942 & 14 40 38.3 +03 28 18.2 & \multicolumn{3}{c}{23.1} \\            
        18266 & 14 40 41.9 +03 28 04.4 & \multicolumn{3}{c}{ 35.6}\\    
        18850 & 14 40 41.9 +03 28 04.4 & \multicolumn{3}{c}{ 39.6}\\   
 \hline 
 \end{tabular}
 \end{table*}

\begin{figure*}
\centering
\subfloat{\includegraphics[width=75mm]{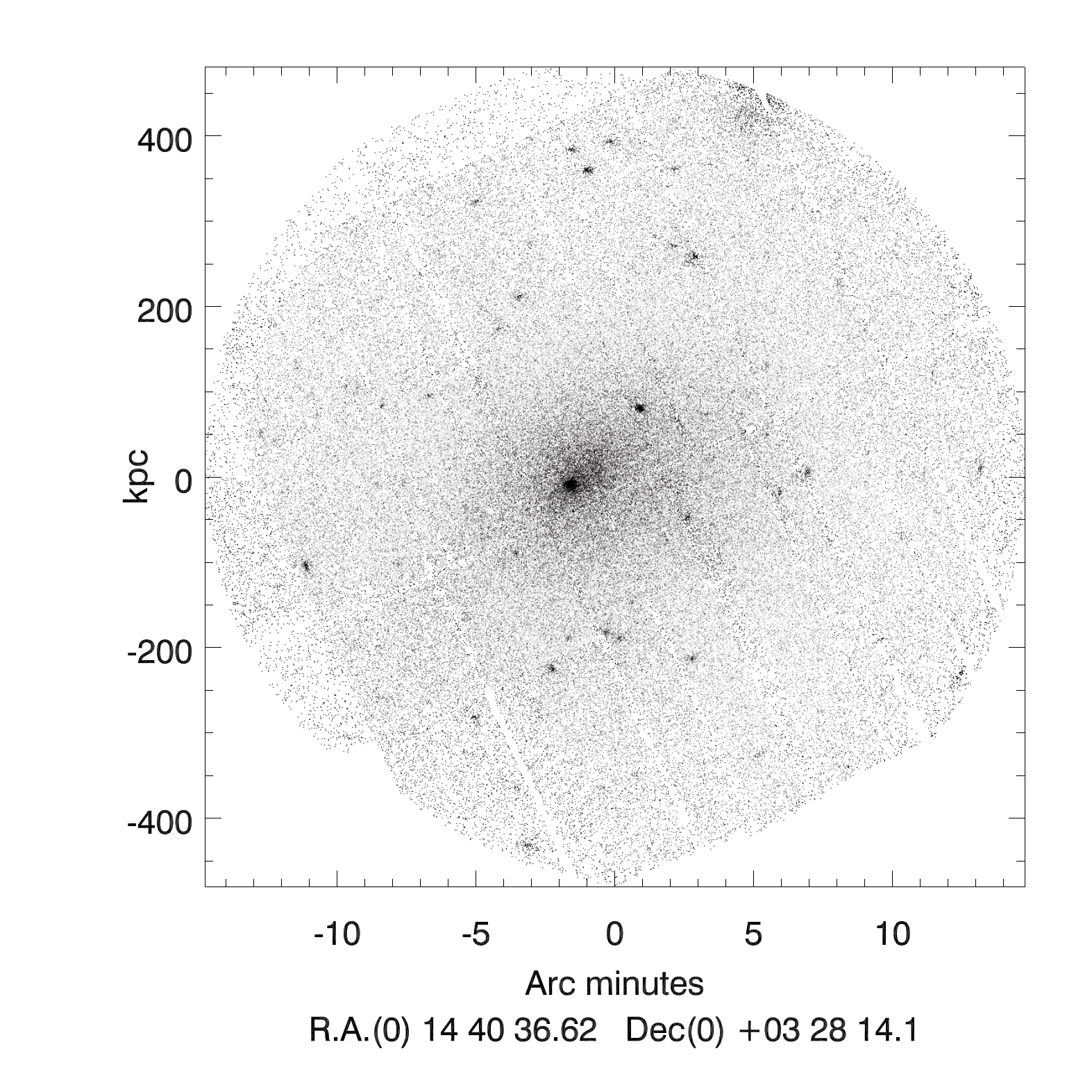}}
\subfloat{\includegraphics[width=75mm]{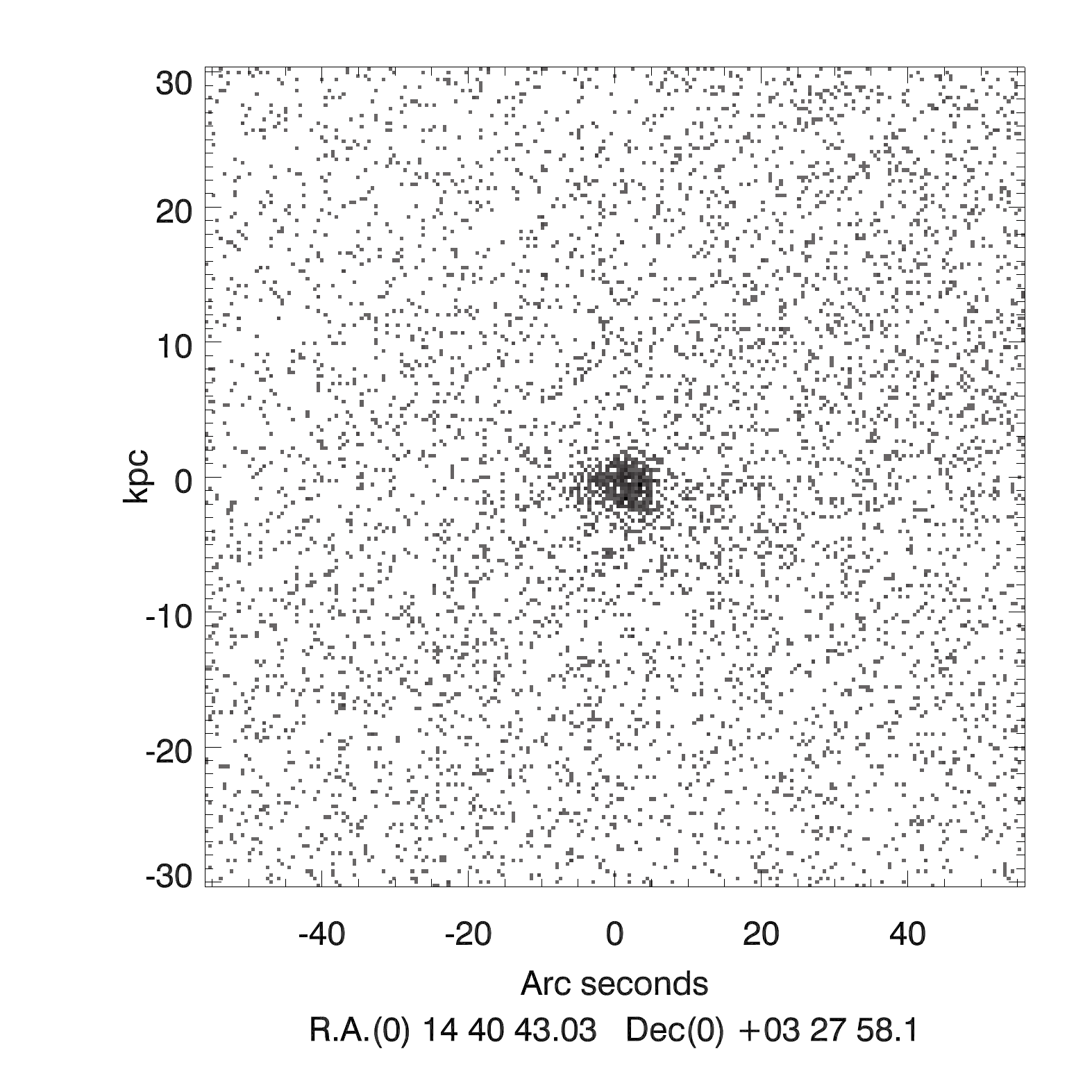}}
\caption{Background subtracted, exposure corrected photon images of \textit{XMM-Newton} (\textit{left}) and \textit{Chandra} (\textit{right}) in soft X-ray (0.3-2.5 keV) band. Although a mosaic of three \textit{Chandra} observations for background estimations were used, an enlarged image of \textit{Chandra} (\textit{right}) is shown in order to emphasize the BCG emission.}
\label{fig:photon}
\end{figure*}

In this work, we used archival \textit{XMM-Newton} EPIC observations (PN, MOS1, and MOS2) for one pointing and three \textit{Chandra} ACIS-I observations for our analysis of the X-ray properties of MKW 08. The specifications of the data are summarized in Table~\ref{tab:obslog}. In Fig.~\ref{fig:photon}, we present the background subtracted, exposure corrected photon images of the \textit{XMM-Newton} and \textit{Chandra} observations.  

\subsubsection{The \textit{XMM-Newton} dataset}

The \textit{XMM-Newton} archival data (observation ID: 0300210701) were primarily used to investigate the ICM surface brightness and thermal structure on large, cluster-size spatial scales, up to \textit{r} $\simeq$ 12.5$\arcmin$ ($\approx$ 420 kpc). Instead, for the spectral analysis of the NGC 5718, we selected circular region with \textit{r} $\simeq$ 10 kpc centered at the BCG.

\subsubsection{The \textit{Chandra} dataset}

The archival \textit{Chandra} ACIS-I observations (observation IDs: 4942, 18266, 18850) were used to investigate the central region of the cluster in detail with a higher angular resolution by studying the surface brightness profile within \textit{r} $\simeq$ 4$\arcmin$ ($\approx$ 130 kpc). In addition, a central circular region with  \textit{r} = 3 kpc was selected to analyze the emission from NGC 5718 corona and its central AGN.

\subsection{Data reduction}
\label{sec:datapreparation}

For both of the \textit{XMM-Newton} and \textit{Chandra} observations, photo detection events are filtered from soft proton flares via a temporal wavelet filtering of their "soft" (1-5 keV) and "hard" (10-12 keV) light curves. Composite event-lists are subsequently rebinned in 3-dimensional position-energy cubes that match the angular and spectral resolution of each instrument. For each element of the data cubes, we associate an effective exposure and a background noise value. 

As detailed in \citet{Bourdin08}, effective exposure values take the spectral and spatial dependences of the mirror effective areas as well as filter absorptions and detector quantum efficiencies into account, and combine them with CCD gaps and bad pixels positions. These components are extracted from the calibration data bases used to pre-process the event lists, namely the CCF and the CALDB 4.3.3 for \textit{XMM-Newton} and \textit{Chandra}, respectively. Point sources are identified from a wavelet analysis and masked for any analysis of extended emissions, including the ICM and unresolved sky background. An exception to this is the BCG, since it is one of the main interests of this study.

\begin{figure}
\resizebox{\hsize}{!}{\includegraphics{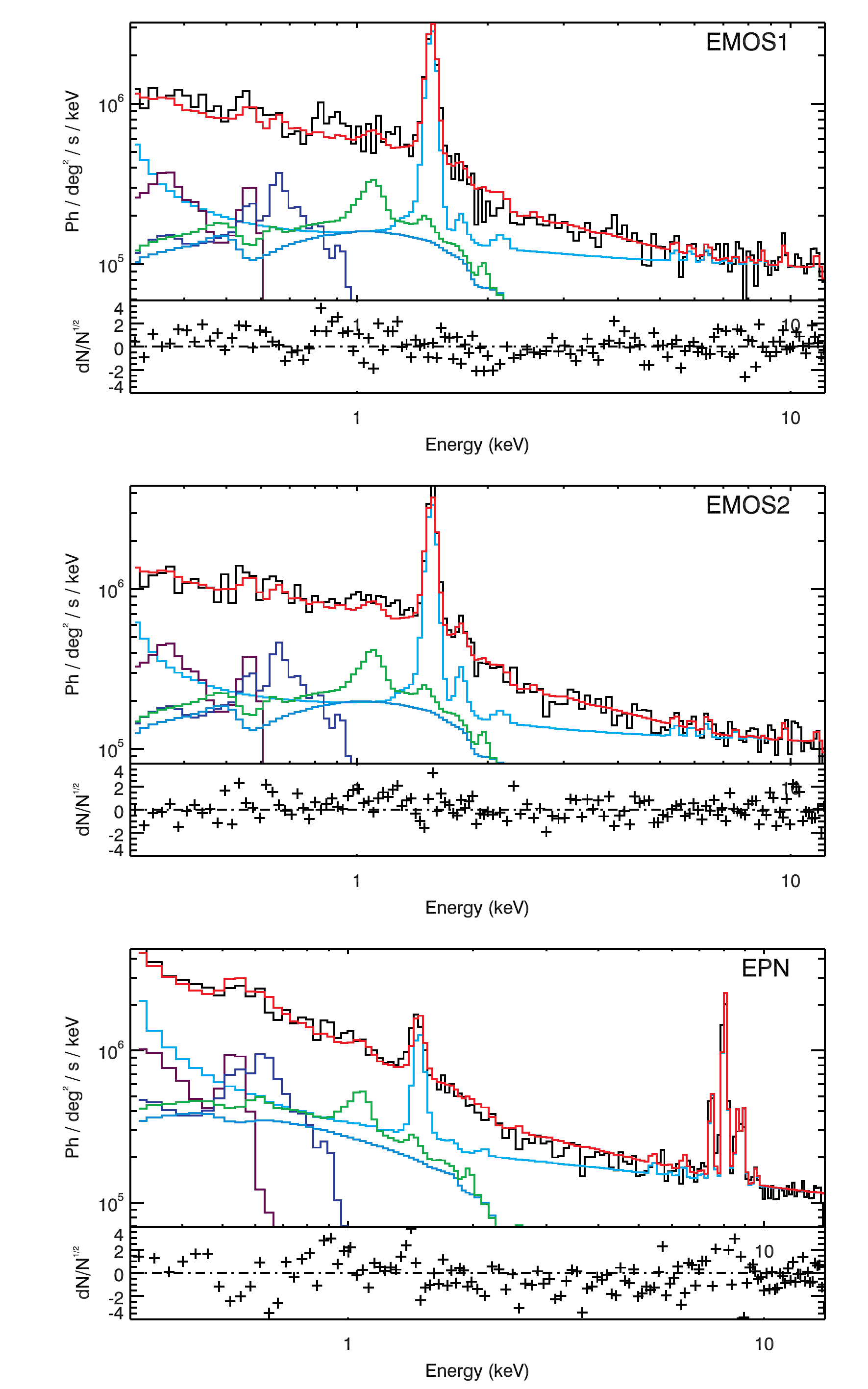}}
\caption{Joint-fit of \textit{XMM-Newton} EPIC background and cluster emission from MKW 08. \textit{From top to bottom}: background fitting of MOS1, MOS2 and PN cameras.  Total number of counts (\textit{black}); thermal emission of LHB (\textit{purple}); thermal emission of GTE (\textit{dark blue}); power-law emission from CXB (\textit{blue}); quiescent particle background emission (\textit{cyan}); cluster emission (\textit{green}); total background emission (\textit{red}). Residuals correspond to the total photon count with respect to the ICM and total background emission.}
\label{fig:XMMback}
\end{figure}

Extended background components can be divided into two main types: astrophysical X-ray background and nonastrophysical X-ray background. Astrophysical X-ray background components are: Galactic thermal emission (GTE), which is the thermal X-ray emission coming from the halo of the Milky Way modeled with two absorbed thermal components $kT_{1}$=0.099 keV and $kT_{2}$=0.248 keV \citep{Kuntz00}; local hot bubble (LHB), an unabsorbed thermal component kT=0.1 keV \citep{Kuntz00}; cosmic X-ray background (CXB) is due to the contribution of mostly unresolved AGNs as well as hot stars and galaxies and modeled by an absorbed power-law with photon spectral index $\gamma$ =1.42 \citep{Lumb02}. 

\begin{figure}
\resizebox{\hsize}{!}{\includegraphics{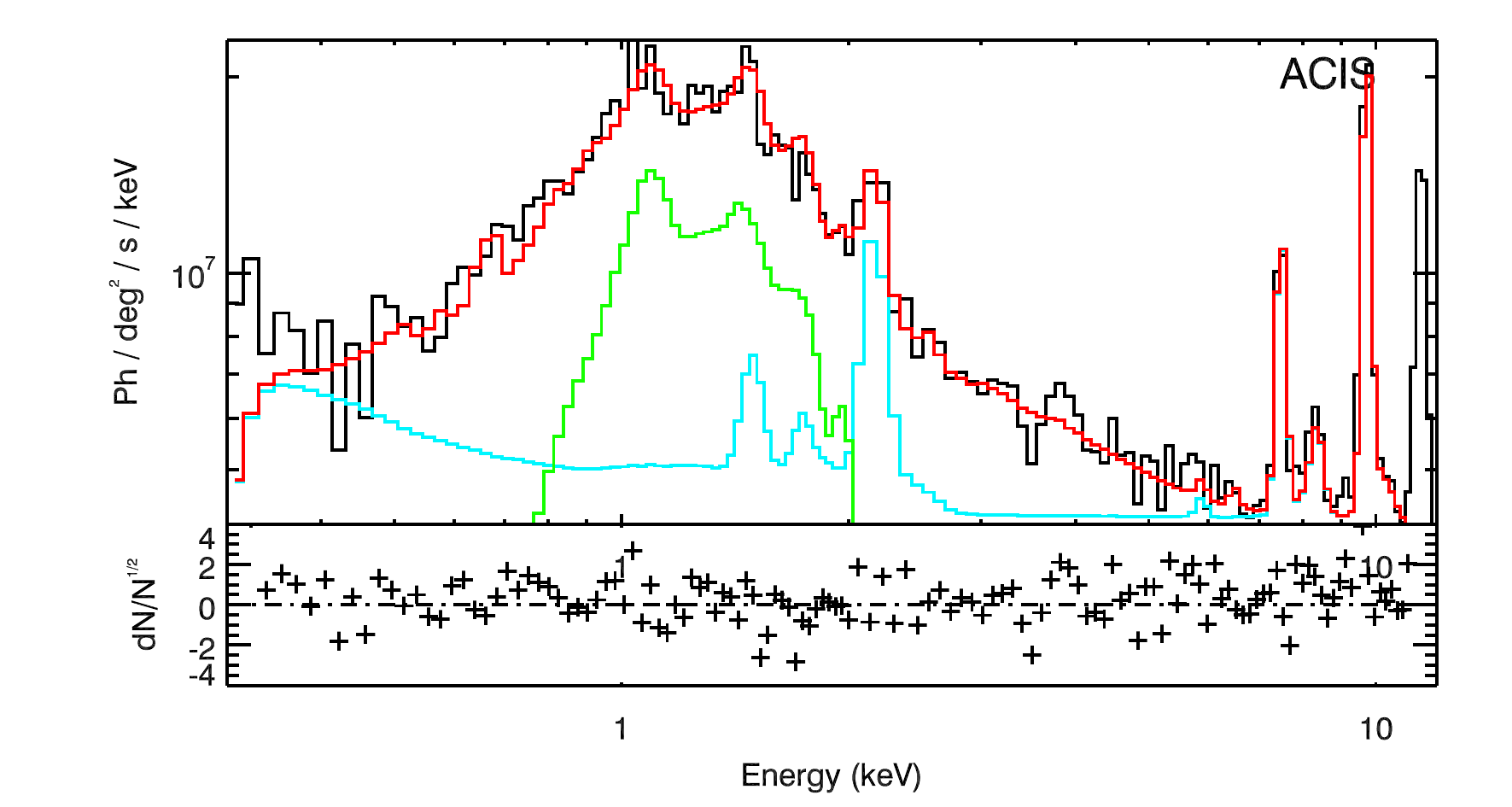}}
\caption{\textit{Chandra} ACIS-I background fitting of three combined observations. Total number of counts (\textit{black}); quiescent particle background emission (\textit{cyan}); cluster emission (\textit{green}); total background emission (\textit{red}). Residuals correspond to the total photon count with respect to the ICM and total background emission.}
\label{fig:Chandraback}
\end{figure}

For modeling the astrophysical X-ray background, the background and cluster emissions are jointly fit. We selected an external annulus for \textit{XMM-Newton} with 12.5$\arcmin$ < \textit{r} < 14.5$\arcmin$, where LHB and GTE are modeled and CXB is fit to the blank-sky values \citep{Read03}. Whereas for \textit{Chandra} observations, we were able to use an annular region of 10$\arcmin$ < \textit{r} < 12$\arcmin$ thanks to the off-set observations. For the astrophysical X-ray background estimation of \textit{Chandra}, we have used the \textit{XMM-Newton} LHB and GTE normalizations of the background components in order to improve the background estimation, since the \textit{XMM-Newton} instruments collect a larger number of photons, which is visible from the photon counts of the background fitting in Fig.~\ref{fig:XMMback} and Fig.~\ref{fig:Chandraback}. For the CXB emission, we have used the blanksky estimates from \citet{Bartalucci14}. We estimated and normalized the sky background components using the \textit{XMM-Newton} EPIC observation, that is characterized at the same time by a more extended field of view and a larger effective area than the \textit{Chandra} ACIS-I observation. 

Figure~\ref{fig:XMMback} and Fig.~\ref{fig:Chandraback} illustrate the background model best-fits for the described regions where the signal to noise ratio (S/N) is at its smallest value, and the residual of the data set with respect to the background and cluster emission. Although we have included the astrophysical X-ray background components for the \textit{Chandra} observations, their contributions cannot be seen in Fig.~\ref{fig:Chandraback} to due the fact that their surface brightness does not reach the level of the quiescent particle background. Therefore, fluctuations of the sky background component are not expected to significantly affect further spectral analyses, especially considering the fact that sky regions of interest for our study are characterized by even higher S/N values than sky regions chosen for spectral analyses of the background components.

Nonastrophysical X-ray background components can be divided into two components. Quiescent Particle Background (QPB) \citep{Kuntz08} is due to physical particles interacting with the detectors even when the cameras are in closed mode. These particles produce X-ray emission, which is the sum of a continuum emission and fluorescent lines. These elemental fluorescent lines are modeled as Gaussians and cause different responses from \textit{XMM-Newton} MOS and PN cameras. Soft Proton background (SP) is modeled as a broken power-law in the case of a soft flare excess after the preliminary data cleaning. However, none of the observations required an additional power-law component for the soft flare excess. QPB background is treated with analytical approaches explained in Sect.~\ref{sec:xmmparticle} and Sect.~\ref{sec:chandraparticle}.

\subsubsection{Analytic particle background model for \textit{XMM-Newton} EPIC}
\label{sec:xmmparticle}

The analytic particle background model for \textit{XMM-Newton} is deduced from the Filter Wheel Closed (FWC) observation, which is the sum of a set of fluorescence emission lines and a continuum emission. These lines have different responses for EPIC cameras and the continuum emission has a flat response across the detectors. In this model, a set of florescence emission lines are convolved with detector energy responses and added to a quiescent continuum. A product of power-law and an inverted error function increasing in the soft band is used to model the quiescent continuum in order to account for the different spectral shapes in soft and hard energy bands. Line energy values are taken from the values reported from \citet{Lec08} and due to their spatial variations, wavelet filtering of a set of FWC event images in narrow energy intervals around the lines are used to model the emissivity distributions of the prominent line emissions. Further details about the QPB model are provided in by \citet{Bourdin13}. The joint-fit of background and cluster emission of MKW 08, is shown in Fig.~\ref{fig:XMMback}.

\subsubsection{Analytic particle background model for \textit{Chandra} ACIS I}
\label{sec:chandraparticle}

For \textit{Chandra}, we used a particle background model for the estimation of the particle background proposed by \citet{Bartalucci14}. The instrumental background of ACIS-I is composed of fluorescence emission lines and a continuum. In this model, in order to isolate the flux of ACIS-I instrumental background from sky components, an analytical model of the continuum was fit to observations spanning eight years starting from 2001, where the observations were performed in very faint mode along with detectors set in stowed position. Blank-sky observations during this period was used to fit the eleven fluorescence emission lines. This resulting model has been shown to predict the instrumental background with a precision of 2\% in the continuum and 5\% in the lines. More details about spatial variations and parametrization of all components of the model are provided in \citet{Bartalucci14}. The total background and the cluster emission model fitting of our data set is shown in Fig.~\ref{fig:Chandraback}.

\section{Cluster emission}
\label{sec:DataAnalysis}

\subsection{ICM surface brightness and temperature maps}

The ICM surface brightness maps are extracted in the 0.3 - 2.5 keV energy band by correcting photon images for background and effective area. Since such images are corrupted by shot noise, we denoised them by means of a wavelet filtering. Specifically, we extracted a multiscale variance stabilized wavelet transform (MSVST, \citet{Starck09}) of the count rate and background images, then set a threshold for the count rate coefficients as a function of a tabulated confidence level. The denoised ICM surface brightness maps are subsequently reconstructed by subtracting background coefficients to the nonzero coefficients of the count rate maps.

\begin{figure*}
\centering
\subfloat{\includegraphics[width=75mm]{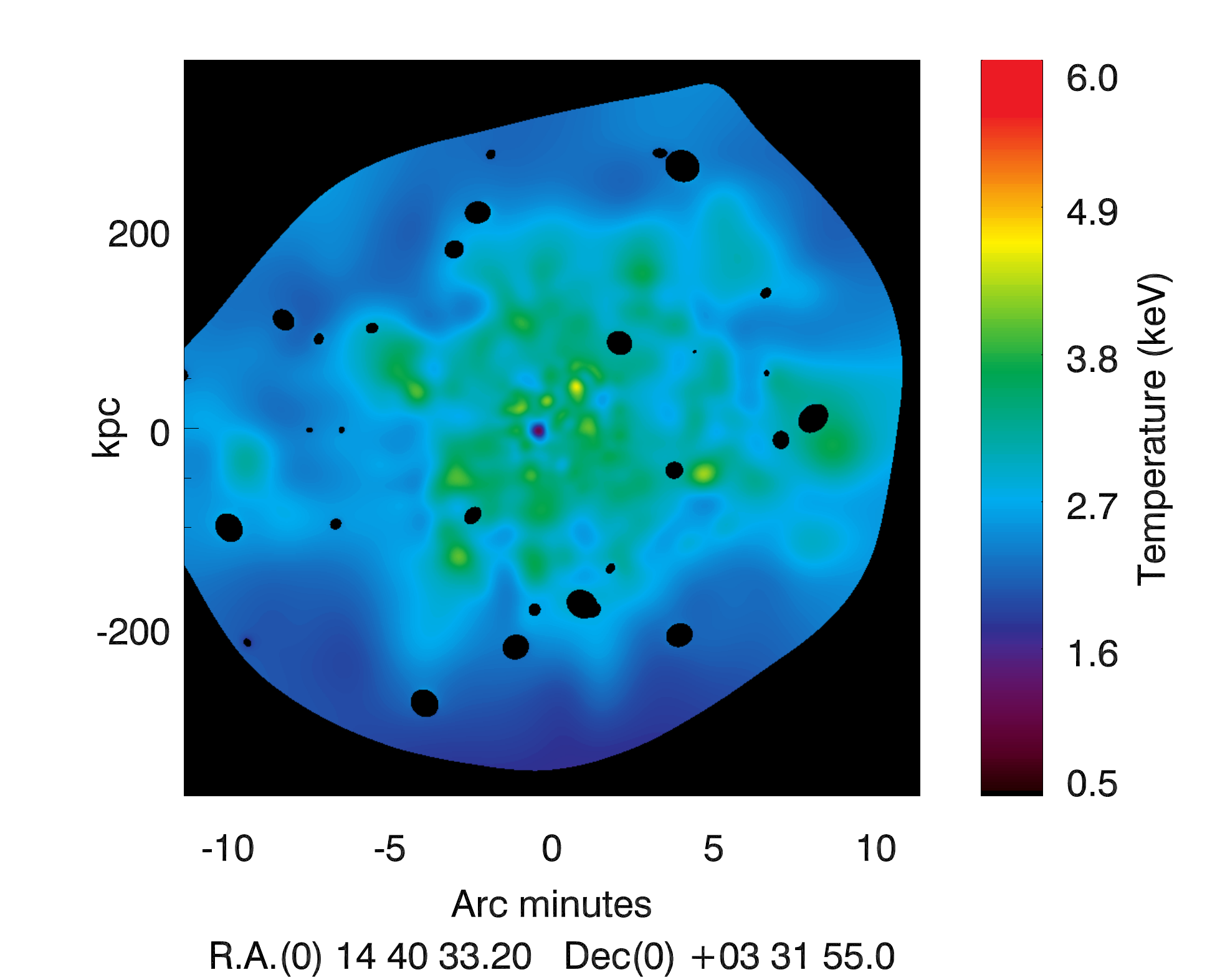}}
\subfloat{\includegraphics[width=59mm]{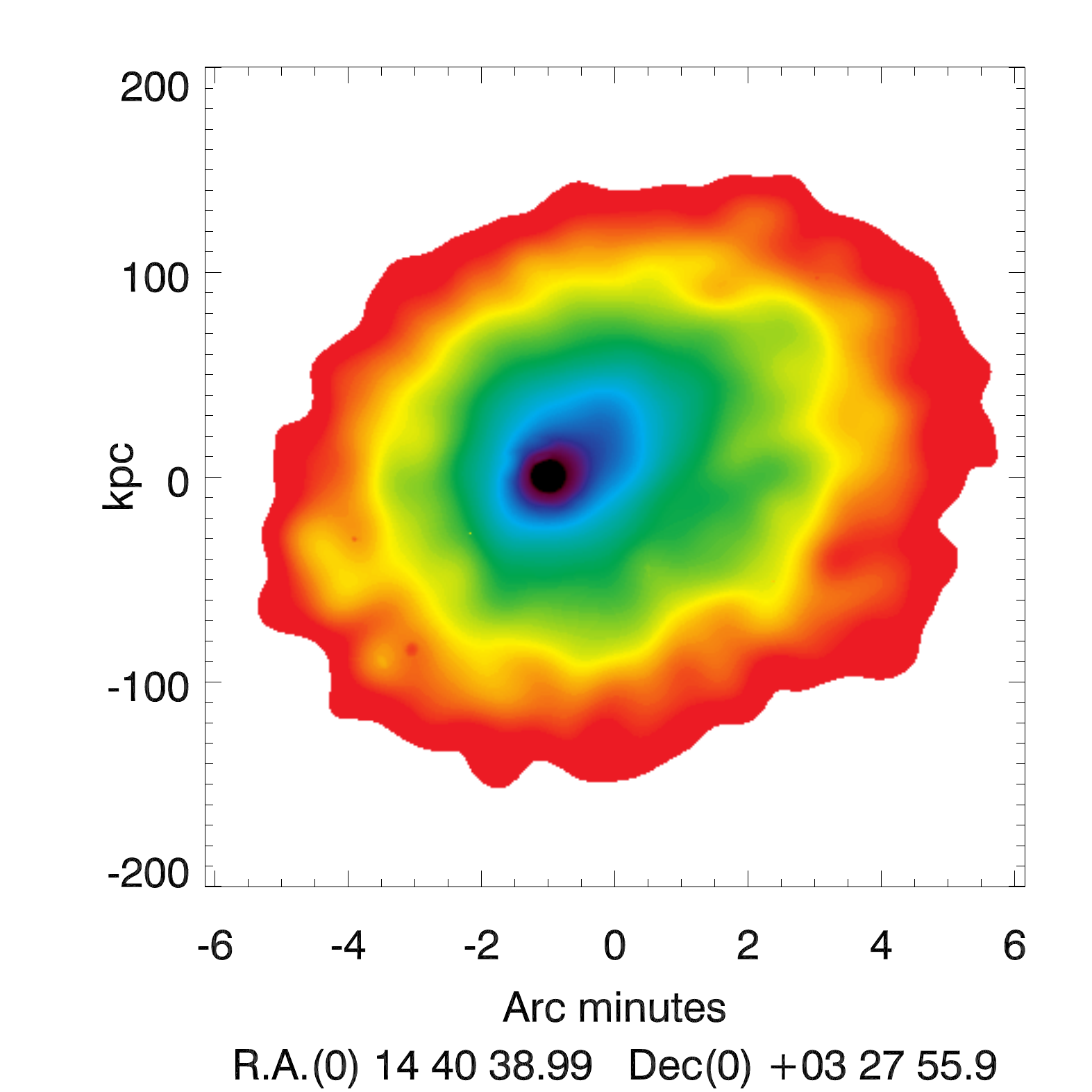}}
\hfill
\subfloat{\includegraphics[width=75mm]{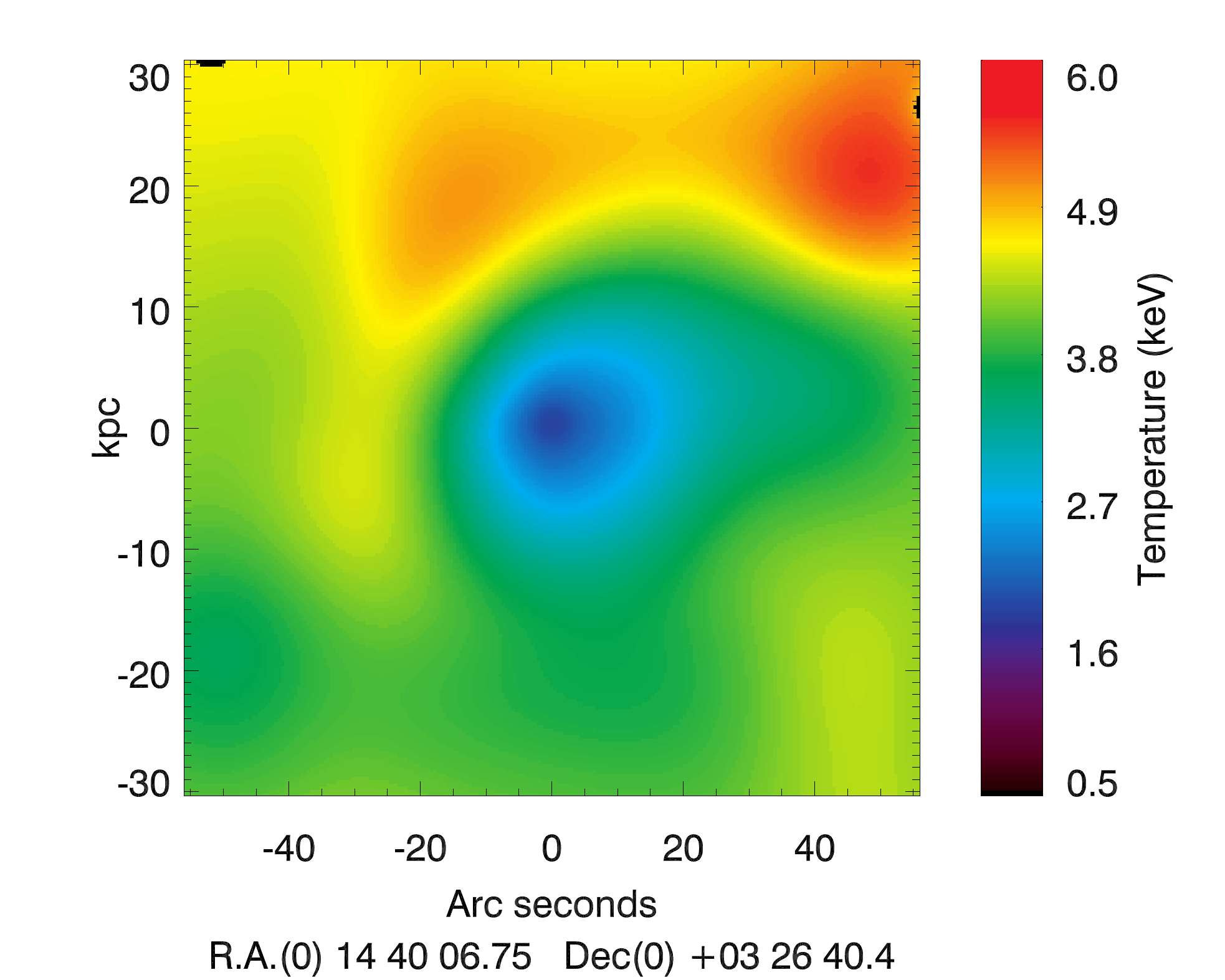}}
\subfloat{\includegraphics[width=59mm]{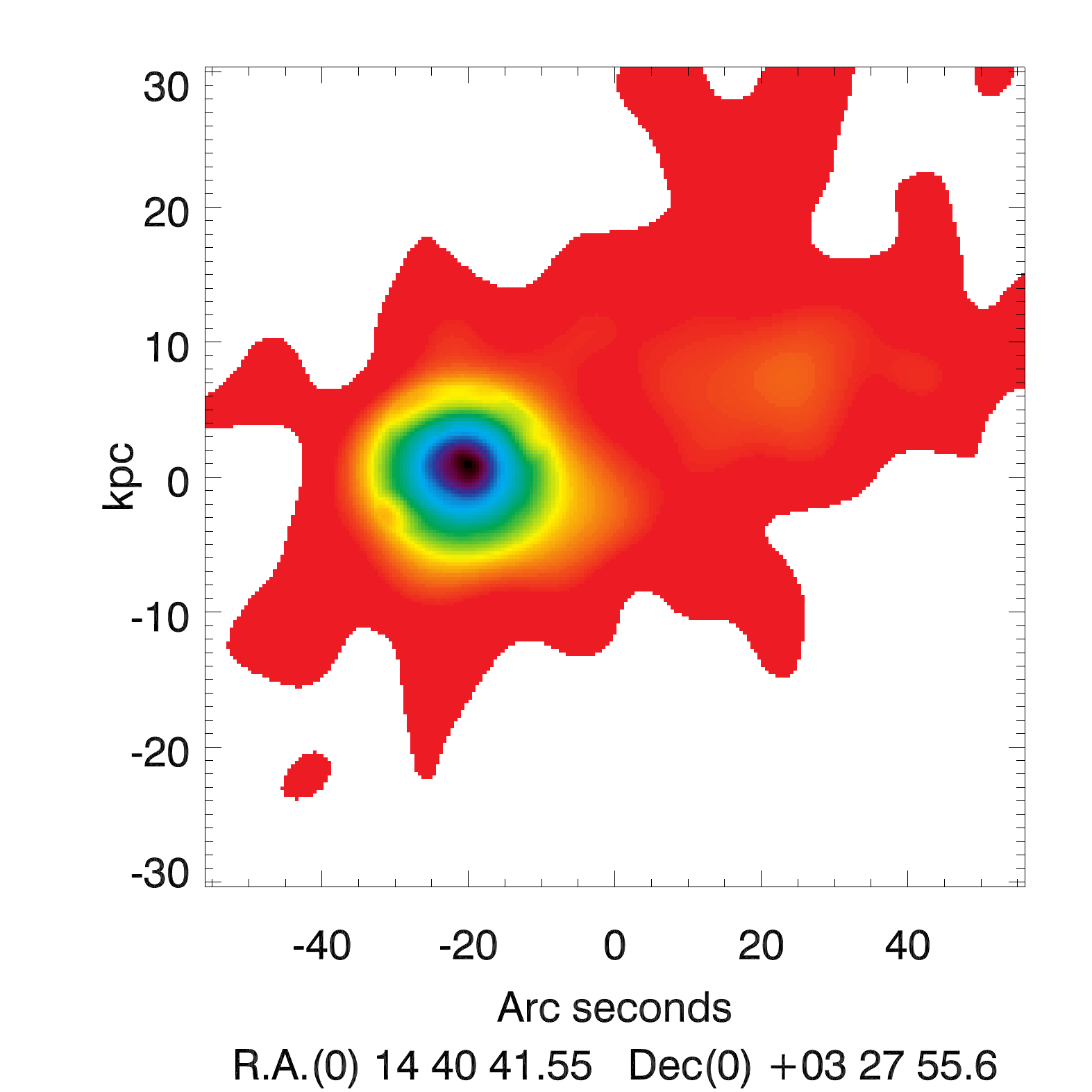}}
\caption{\textit{Upper panel:} \textit{XMM-Newton} temperature map (\textit{left}) and surface brightness map (\textit{right}). \textit{Lower panel:} For the highest angular resolution, \textit{Chandra} temperature map (\textit{left}) and surface brightness map (\textit{right}). Temperature maps are presented within 3$\sigma$  errors. Surface brightness maps are obtained in soft energy band 0.3-2.5 keV within 4$\sigma$ errors. \textit{Chandra} maps are enlarged to emphasize the central region of the cluster.}
\label{fig:XMMtempmap}
\end{figure*}

\begin{figure}
\center
{\includegraphics[width=75mm]{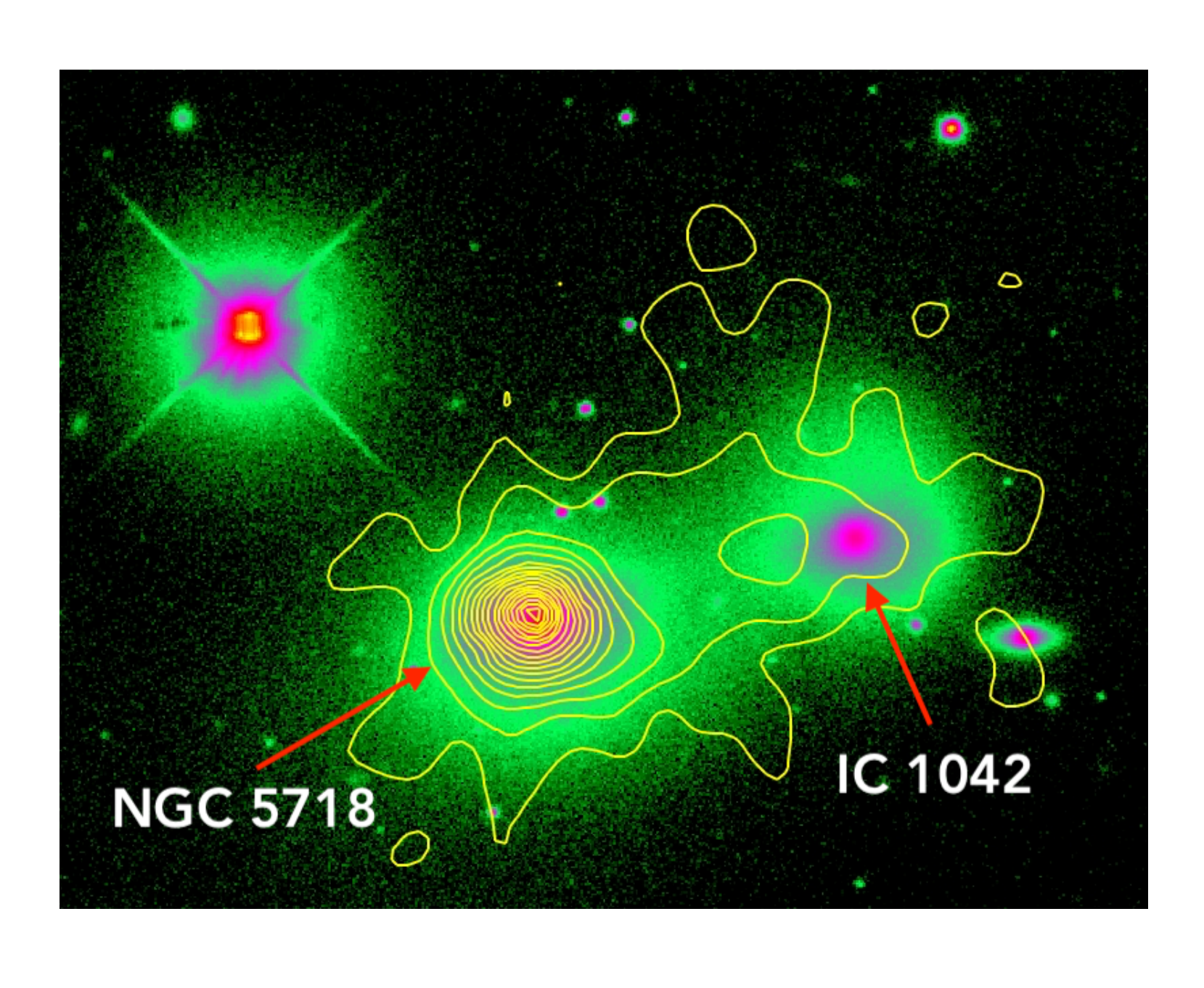}}
\caption{Optical image of the central BCG - NGC 5718 and IC 1042 in the I band overlaid with \textit{Chandra} X-ray contours (\textit{yellow}), which are equispaced by a factor 2$^{1/4}$.}
\label{fig:optical}
\end{figure}

To map the projected ICM temperature, we used the wavelet spectral imaging algorithm introduced in \citet{Bourdin08}. Briefly, spectroscopic temperature values are extracted within overlapping meta-pixels of various sizes, together with their confidence intervals. These temperature values were propagated toward a B2-spline wavelet transform, that is subsequently limited by a given confidence level as a function of the coefficient variance. Local temperatures were estimated via a spectroscopic likelihood maximization, assuming the ICM emission to follow the Astrophysical Plasma Emission Code (APEC, \citet{Smith01}), and the background noise to be modeled as detailed in Sect.~\ref{sec:datapreparation}.

The position of the BCG can be detected on the temperature and surface brightness maps of \textit{XMM-Newton} and \textit{Chandra} in Fig.~\ref{fig:XMMtempmap} due to its low temperature feature, with the guidance of the optical image overlaid with \textit{Chandra} X-ray contours\footnote{The optical image has been retrieved from Sloan Digital Sky Survey III, available via http://www.sdss3.org/dr9/.}.  The \textit{XMM-Newton} temperature map reveals no strong anisotropy on scales larger than \textit{r} $\sim$ 100 kpc, which is typical of a rather relaxed cluster, showing a temperature gradient from the core to the outskirts. The surface brightness map of \textit{XMM-Newton} reveals an elongated cluster core oriented $\sim$ 45 degrees on southeast - northwest axis. The \textit{Chandra} surface brightness map reveals a $\sim$ 40 kpc tail behind the galaxy, however the major axis of the core and the direction of the tail do not coincide. The temperature maps point to a cold reservoir centered at the BCG, which is embedded in the surrounding hotter ICM. The temperature maps are centered at the position of the BCG.

\subsection{X-ray brightness measurements and ICM temperature profiles}
\label{sec:brightness}

The ICM surface brightness and temperature profiles are extracted from a circular region of  \textit{r} = 12.5$\arcmin$ for \textit{XMM-Newton} and  \textit{r} = 4$\arcmin$ for \textit{Chandra}. Both surface brightness profiles were extracted with logarithmic binning of the data in 25 concentric annuli. For the temperature profiles, we used eleven and ten bins for \textit{XMM-Newton} and \textit{Chandra}, respectively.

\begin{figure}
\center
{\includegraphics[width=85mm]{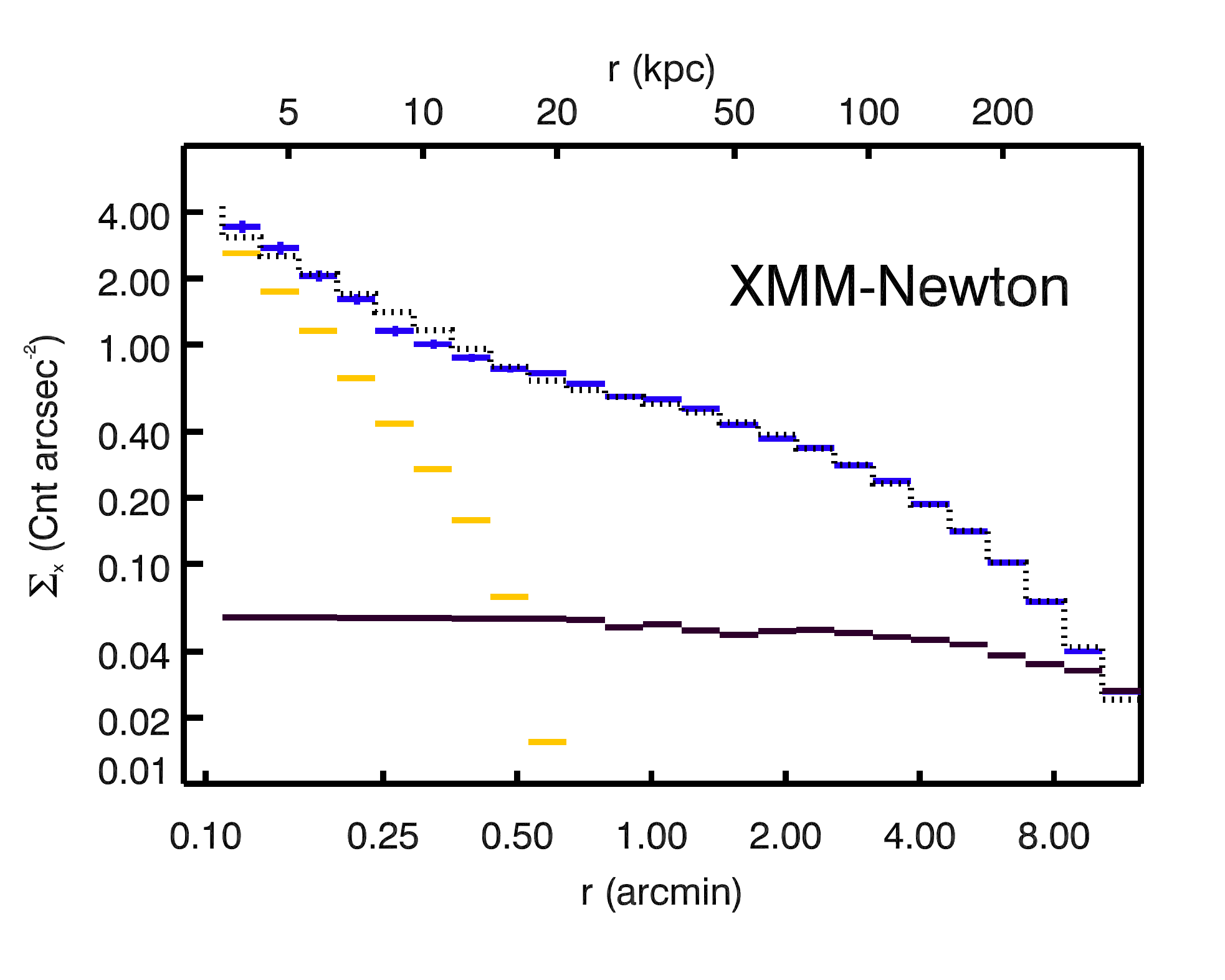}}
\caption{\textit{XMM-Newton} surface brightness fitting with double $\beta$-model in 0.5 - 2.5 keV energy band in circular region with \textit{r} = 12.5$\arcmin$. Background (\textit{black}); \textit{XMM-Newton} point spread function model (yellow); surface brightness of the cluster (\textit{blue}); the double $\beta$-model fit of the projected surface brightness (\textit{dotted black curve}).}
\label{fig:XMMbrightness}
\end{figure}

Assuming spherical symmetry and hydrostatic equilibrium, we employed the modified analytical $\beta$-model described in \citet{Vik06} for the deprojection of  gas density distributions. As described in Eqn.~\ref{eqn:3dGasDensity}, this parametric model (hereafter double $\beta$-model), adds two terms. The first term is a modified beta model that we prevalently used to fit the ICM surface brightness. The second term is an additional beta model that we used to fit the surface brightness of the coronal gas emission from the BCG. Furthermore, the double $\beta$-model includes an $\alpha$ parameter as a fine adjustment to represent the power-law characteristics of the gas density in the centers of clusters, which may deviate from a flat core \citep[see, e.g.,][]{Pointecouteau04}. 

 \begin{equation}
 \label{eqn:3dGasDensity}
 \left[n_{e}n_{p}\right](r)=n_0^2\frac{({r}/{r_c})^{-\alpha}}{(1+{r^2/r_c^2})^{3\beta-{\alpha/2}}}\frac{1}{(1+{{r^\gamma }/{r_s^\gamma})^{\varepsilon /\gamma }}}  \\
 +\frac{n_{02}^2}{(1+{r^2 }/{r_{c2}^2})^{3\beta_2}}.
 \end{equation}
  
\begin{figure}
\center
{\includegraphics[width=85mm]{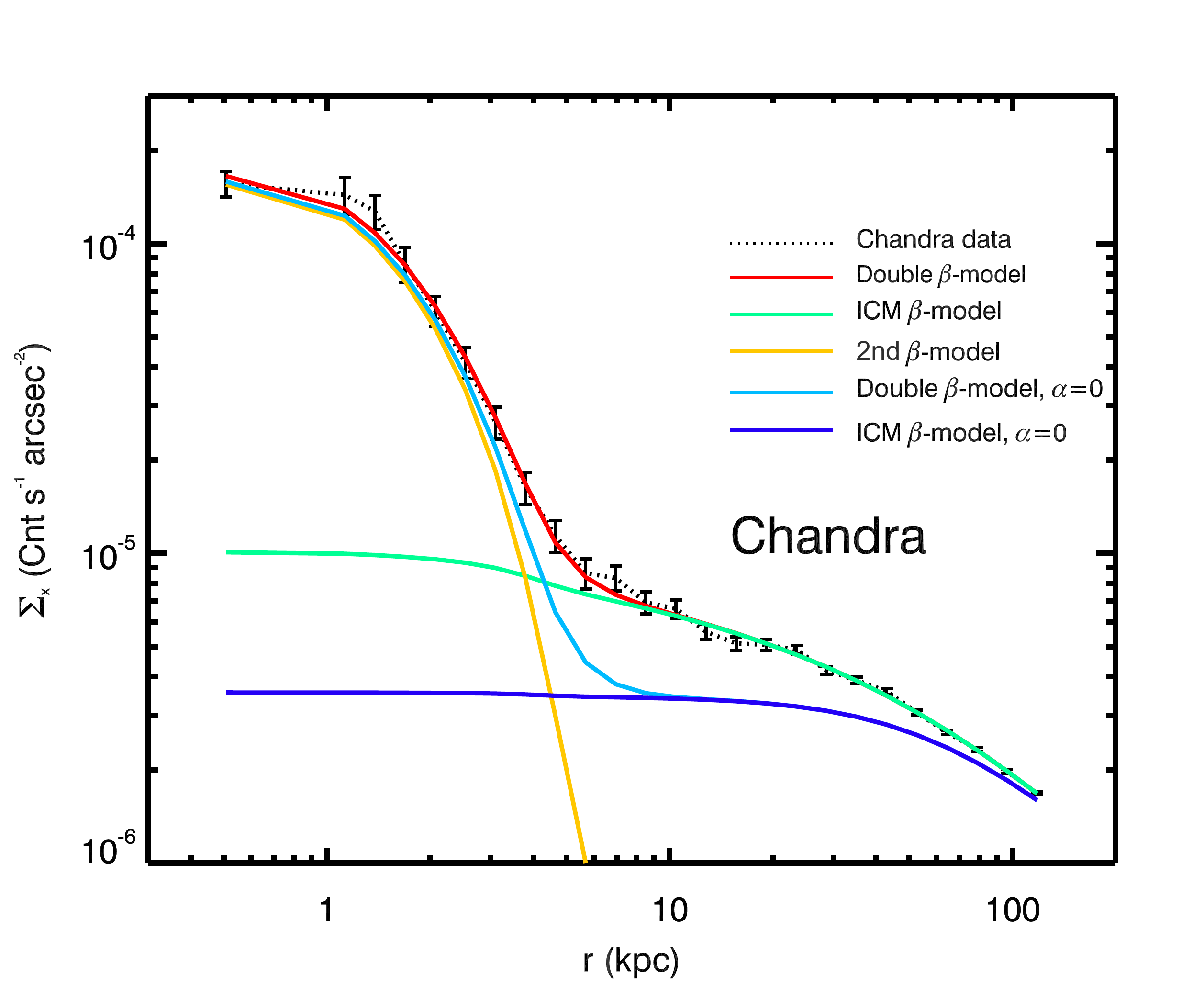}}
\caption{\textit{Chandra} surface brightness fit of double $\beta$-model in 0.5 - 2.5 keV energy band in circular region with \textit{r} = 4$\arcmin$. Projected surface brightness (\textit{dotted black curve}); the double $\beta$-model (\textit{red}); the $\beta$-model corresponding to the ICM (\textit{green}); the second $\beta$-model corresponding to the corona (\textit{yellow}); the double $\beta$-model with $\alpha$ parameter is set to zero (\textit{light blue}); the $\beta$-model of the ICM with $\alpha$ parameter is set to zero (\textit{dark blue}).}
\label{fig:Chandrabrightness}
\end{figure}

By using the proton (\textit{n}$_p$) and electron (\textit{n}$_e$) densities obtained from the fit and integrating the ICM density along the line of sight in the energy band between 0.5 - 2.5 keV, we obtained the projected surface brightness profiles, $\Sigma$$_{x}$(\textit{r}). The surface brightness profiles of \textit{XMM-Newton} and \textit{Chandra} are given in Fig.~\ref{fig:XMMbrightness} and Fig.~\ref{fig:Chandrabrightness} respectively.

For modeling 3-dimensional temperature profiles, we used a five parameter broken power-law with transition region as given in Eqn.~\ref{temp} \citep{Vik06}. Subsequently, we applied an integration along the line of sight to deduce the projected "spectroscopic-like" temperature profiles as described by \citet{Mazzotta04}. 

 \begin{equation}
 \label{temp}
 T_{3D}(r)=T_{0}\frac{({r}/{r_t})^{-a}}{(1+(r/r_t)^{b})^{c /b}}.
 \end{equation}

Projected temperature profiles plotted over the corresponding 3-dimensional temperature profiles are shown in Fig.~\ref{fig:ktprof}. The corresponding count rates are given in Table~\ref{tab:Countrates} in the appendix section of our work. These methods are described in detail by \citet{Bourdin08}.

As shown by the dotted curve in Fig.~\ref{fig:XMMbrightness}, the ICM surface brightness is ideally fit by the projection of an analytical electron density profile given by Eqn.~\ref{eqn:3dGasDensity} from a few kpc to about $\approx$ 400 kpc. This profile is plotted together with a radial profile of the \textit{XMM-Newton} aim point Point Spread Function (PSF). The surface brightness profile is clearly shallower than the PSF profile, which demonstrates that \textit{XMM-Newton} allows us to spatially resolve an extended X-ray emission around the BCG. 

\begin{figure}
\centering
{\includegraphics[width=75mm]{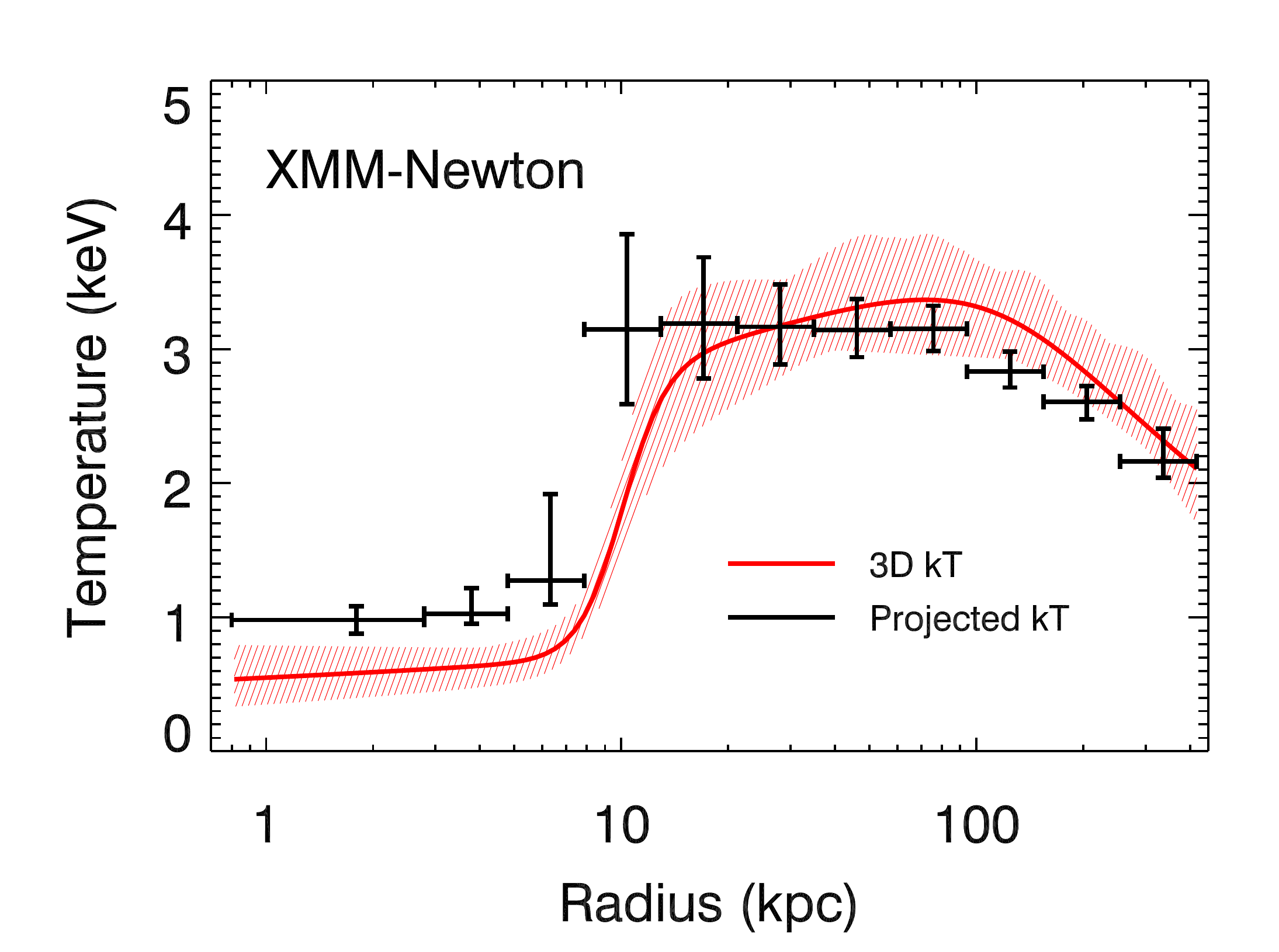}}
{\includegraphics[width=75mm]{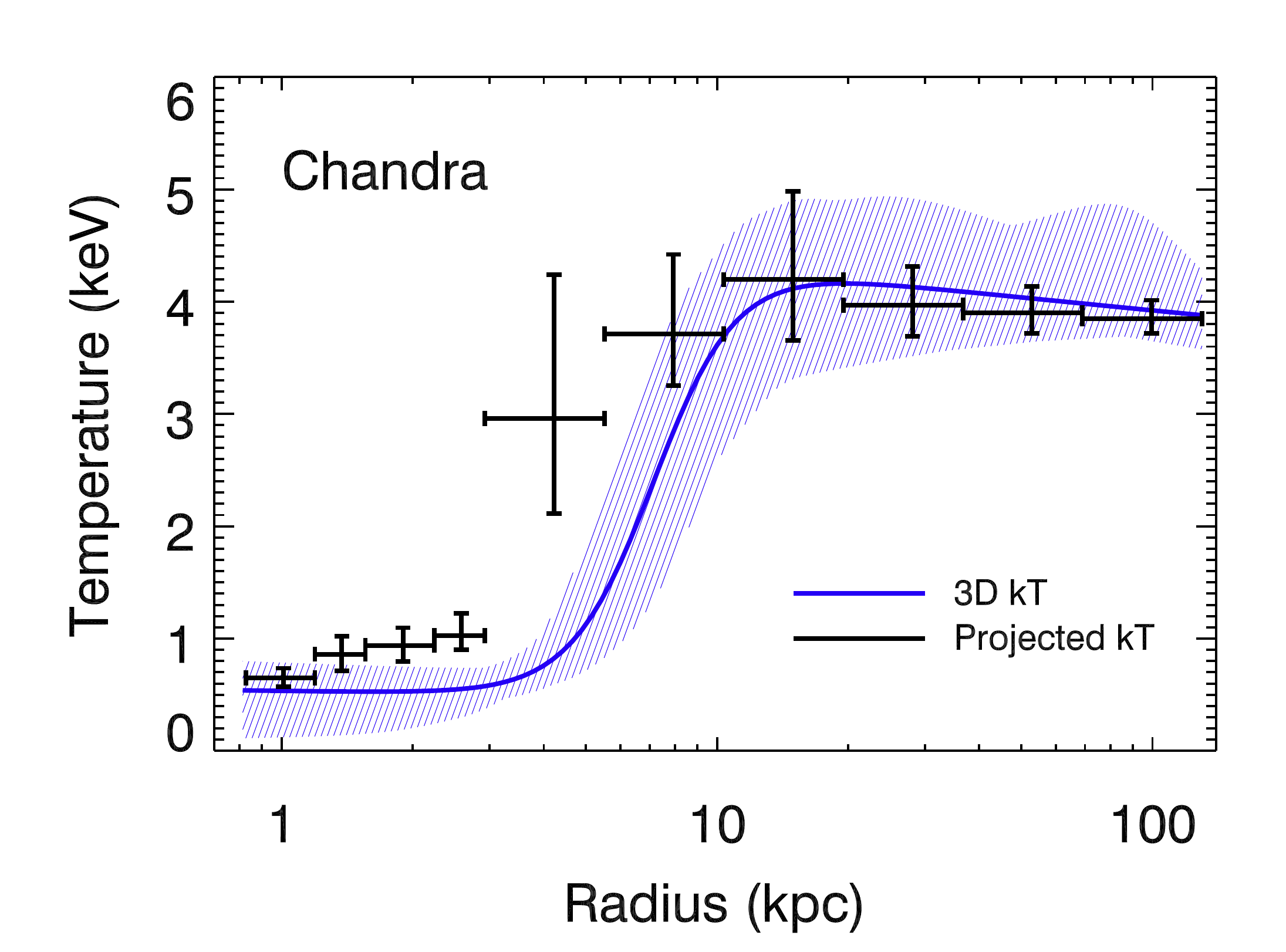}}
\caption{Projected radial ICM temperature values plotted over the deprojected temperature profiles corresponding to \textit{r} = 12.5$\arcmin$ for \textit{XMM-Newton} (\textit{top}) and \textit{r} = 4$\arcmin$ for \textit{Chandra} (\textit{bottom}) as described in Sect.~\ref{sec:brightness}. Confidence envelopes for \textit{XMM-Newton} (\textit{red}) and \textit{Chandra} (\textit{blue}) were propagated from Monte Carlo realizations of the fit analytical functions.}
\label{fig:ktprof}
\end{figure}

The higher \textit{Chandra} angular resolution allows us to better resolve the X-ray surface brightness observed in the neighborhood of the corona. As shown in Fig.~\ref{fig:Chandrabrightness}, the \textit{Chandra} surface brightness profile can be divided in two components that stand for the ICM and coronal gas emissions. In the radii range from 10 kpc to 100 kpc, where the ICM emission dominates, the \textit{Chandra} surface brightness profile is accurately fit by the projection of an electron density profile that follows a single $\beta$-model corrected by a power-law. The parametric form of such a density profile is given by the first fraction in the right-hand side of Eqn.~\ref{eqn:3dGasDensity}. 

In order to visualize the individual contributions of these terms to the total model, we disentangled the components of the double $\beta$-model applied to \textit{Chandra}. Firstly, we set the ICM emission normalization, \textit{n$_0$}, to zero in order to obtain the contribution from the coronal gas emission. Then, retracting ICM normalization back to its original value and setting the coronal gas emission normalization, \textit{n$_{02}$} to zero, we acquired the ICM emission curve. These curves are shown in Fig.~\ref{fig:Chandrabrightness} in yellow and green respectively. These curves intersect at a location \textit{r} $\approx$ 4 kpc, within which the coronal gas emission is dominant. The region enclosed by these two curves and the red curve (the double $\beta$-model), points to the existence of an interface region. 

In order to illustrate the fact that the ICM density profile is more cuspy than a simple beta-model in the cluster center, we set only the $\alpha$ parameter to zero and obtained the light blue curve in Fig.~\ref{fig:Chandrabrightness}. This curve shows that $\alpha$ is almost exclusively a feature of the ICM emission. Keeping $\alpha$ parameter at zero and furthermore, setting \textit{n$_{02}$} to zero, the resulting dark blue curve shows that the power-law is needed to fit a surface brightness excess of about a factor two at \textit{r} $\approx$ 10 kpc with respect to the projection of a single $\beta$-model that would appear flat inside its characteristic radius. Interior to \textit{r} $\approx$ 4 kpc, where emission originating from the coronal gas dominates, a secondary $\beta$-model must be added to the ICM component. Interestingly, this parametric model suggests that both ICM and coronal gas contribute to the observed surface brightness in the radii range of 4 - 8 kpc, which let us infer the existence of a smooth interface region separating the two media.

The projected temperature profile of \textit{XMM-Newton} as shown in Fig.~\ref{fig:ktprof} reveals an isothermal region within \textit{r} $\leqslant$ 5 kpc with a temperature \textit{kT} $\simeq$ 1 keV. However in the range of 5 kpc $\leqslant$ \textit{r} $\leqslant$ 10 kpc, the temperature values show significant uncertainties, likely due to the interface of the corona and the ICM. Inside this region, the deprojected temperature profile points to a sharp increase in temperature values, followed by a gradual increase up to \textit{r} $\simeq$ 80 kpc, then steadily decreases to \textit{kT} $\simeq$ 2 keV around \textit{r} $\simeq$ 400 kpc. 

The \textit{Chandra} projected temperature values show a gradient inside \textit{r} $\leqslant$ 3 kpc. However, the deprojection curve shows the gas is isothermal up to \textit{r} $\simeq$ 4 kpc therefore the gradient is not physical but is due to the projection. With its superior angular resolution, \textit{Chandra} reveals that the isothermal region is enclosed only inside \textit{r} $\leqslant$ 4 kpc.

\begin{figure}
\centering
\includegraphics[width=80mm]{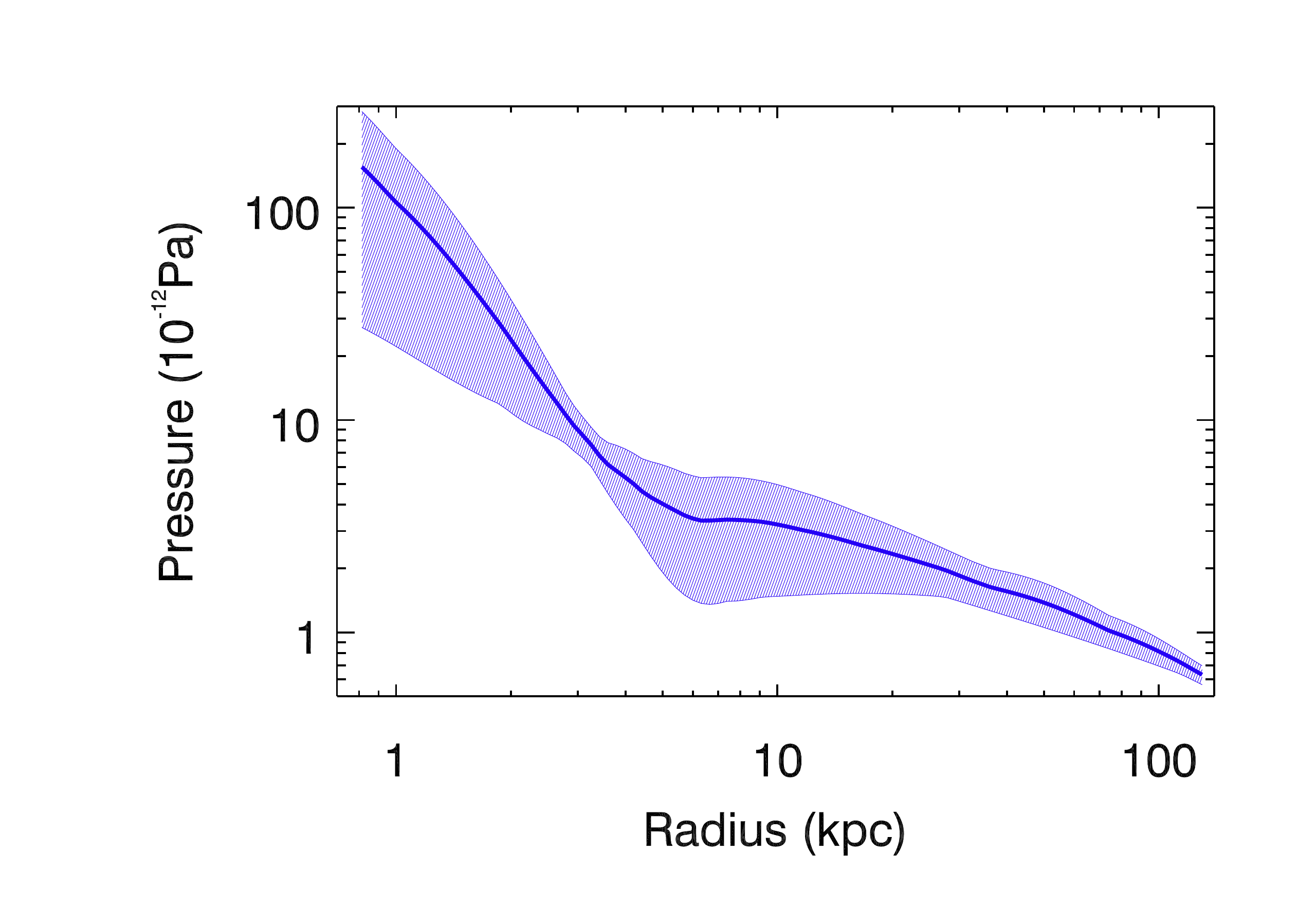}
\caption{Hot gas pressure profile centered at BCG.}
\label{fig:SP}
\end{figure}

As we have used the \textit{XMM-Newton} data for projected surface brightness and temperature profile analysis for the ICM, we turned our focus on the \textit{Chandra} profiles to investigate the interface region between the corona and the ICM. Using the analytic density and 3-dimensional temperature profiles shown in Fig.~\ref{fig:Chandrabrightness} and Fig.~\ref{fig:ktprof}, we derived the 3-dimensional pressure profile shown in Fig.~\ref{fig:SP} using fundamental thermodynamic equations. Moreover, with the minimization of the distance between the projected models and Monte Carlo simulations of the observed values, we obtained the confidence intervals of the 3-dimensional profiles. The pressure profile shown in Fig.~\ref{fig:SP} reveals the presence of an interface region again at 4 $\leqslant$ \textit{r} $\leqslant$ 10 kpc.

\section{Spectral analysis of the BCG: NGC 5718}
\label{sec:NGC5718}

For the spectral analysis of NGC 5718 using the \textit{XMM-Newton} data, we primarily focused on a circular region with \textit{r} = 10 kpc ($\approx$18$\arcsec$) centered at the position of the source. This radius encloses a region whose temperature profile exhibits a temperature gradient as shown in Fig.~\ref{fig:ktprof}. In addition, the \textit{Chandra} temperature and pressure profiles suggest that the corona - ICM interface is enclosed inside \textit{r} = 10 kpc ($\approx$18$\arcsec$) as explained in Sect.~\ref{sec:brightness}. We aimed to observe the thermal emission components of coronal gas and the interface region. Following the \textit{XMM-Newton} spectral analysis, we focused on the \textit{r} = 3 kpc region centered at the BCG with the \textit{Chandra} observations in order to isolate the emission from NGC 5718 thanks to the superior angular resolution of \textit{Chandra}. 

In our spectral analysis, we used \textit{XSPEC} v. 12.10.0. Photon counts used in spectra were not grouped, therefore we do not apply the $\chi^{2}$ statistics, which requires a minimum of 25 counts in each energy bin. Since the data follow a Poisson distribution, we applied the maximum likelihood-based statistics (hereafter, C-stat) for Poisson data proposed by \citep{cash79}. All errors on spectral the parameter values in this section is presented with 1$\sigma$ confidence levels, unless otherwise stated.

\subsection{\textit{XMM-Newton} EPIC spectral fitting}
\label{sec:XMMfit}

\begin{figure}
\centering
{\includegraphics[width=70mm]{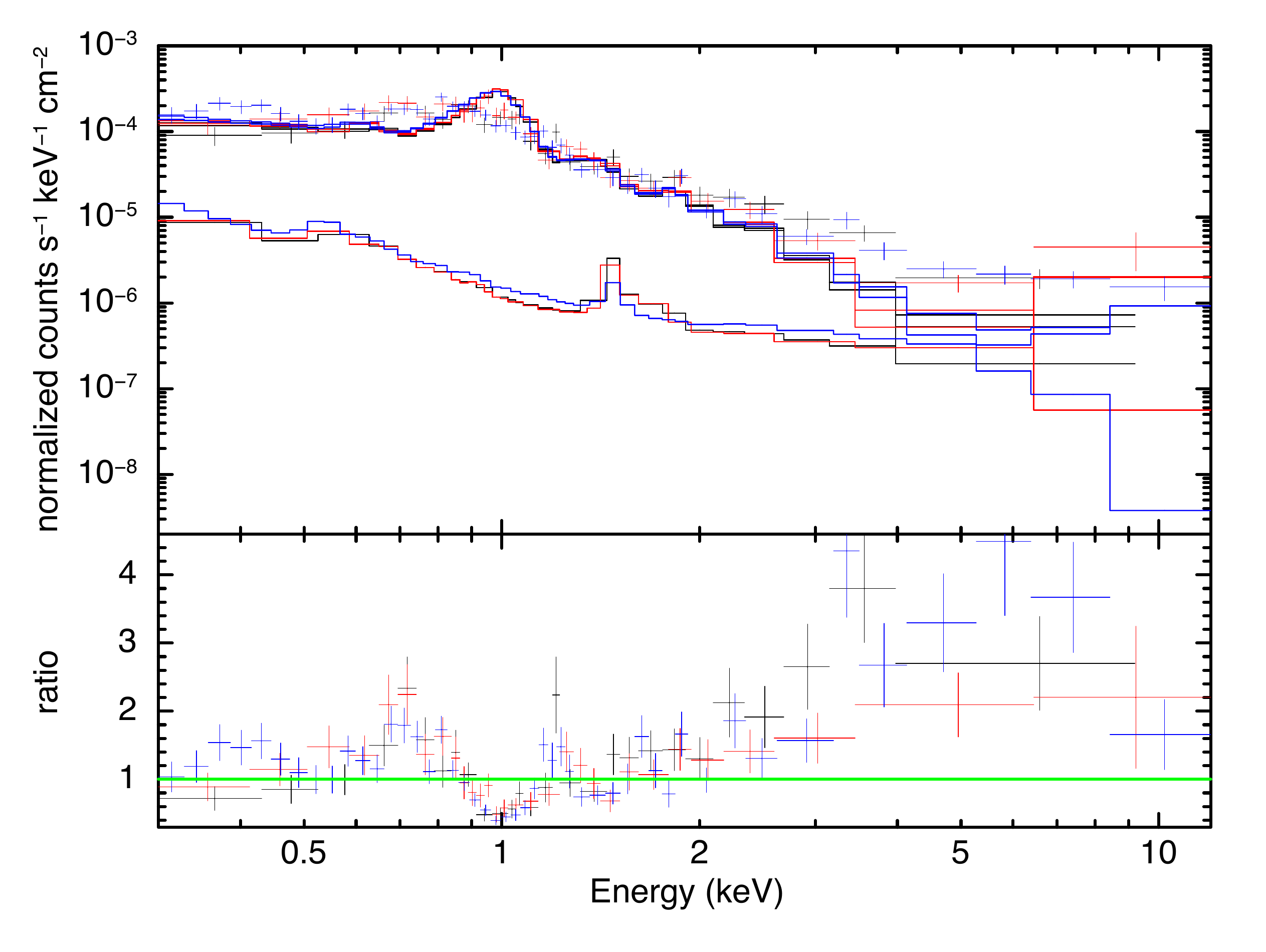}}
{\includegraphics[width=70mm]{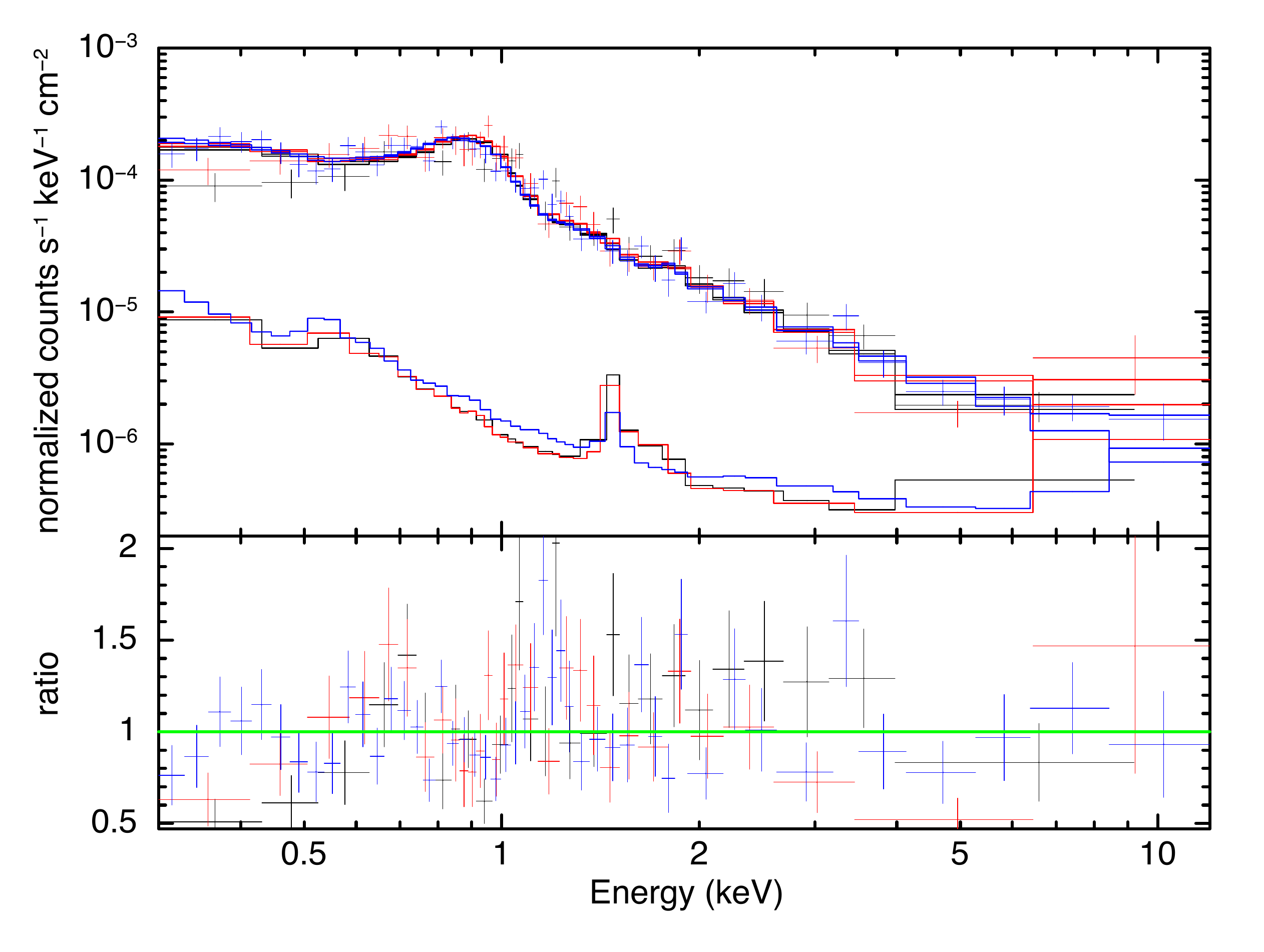}}
{\includegraphics[width=70mm]{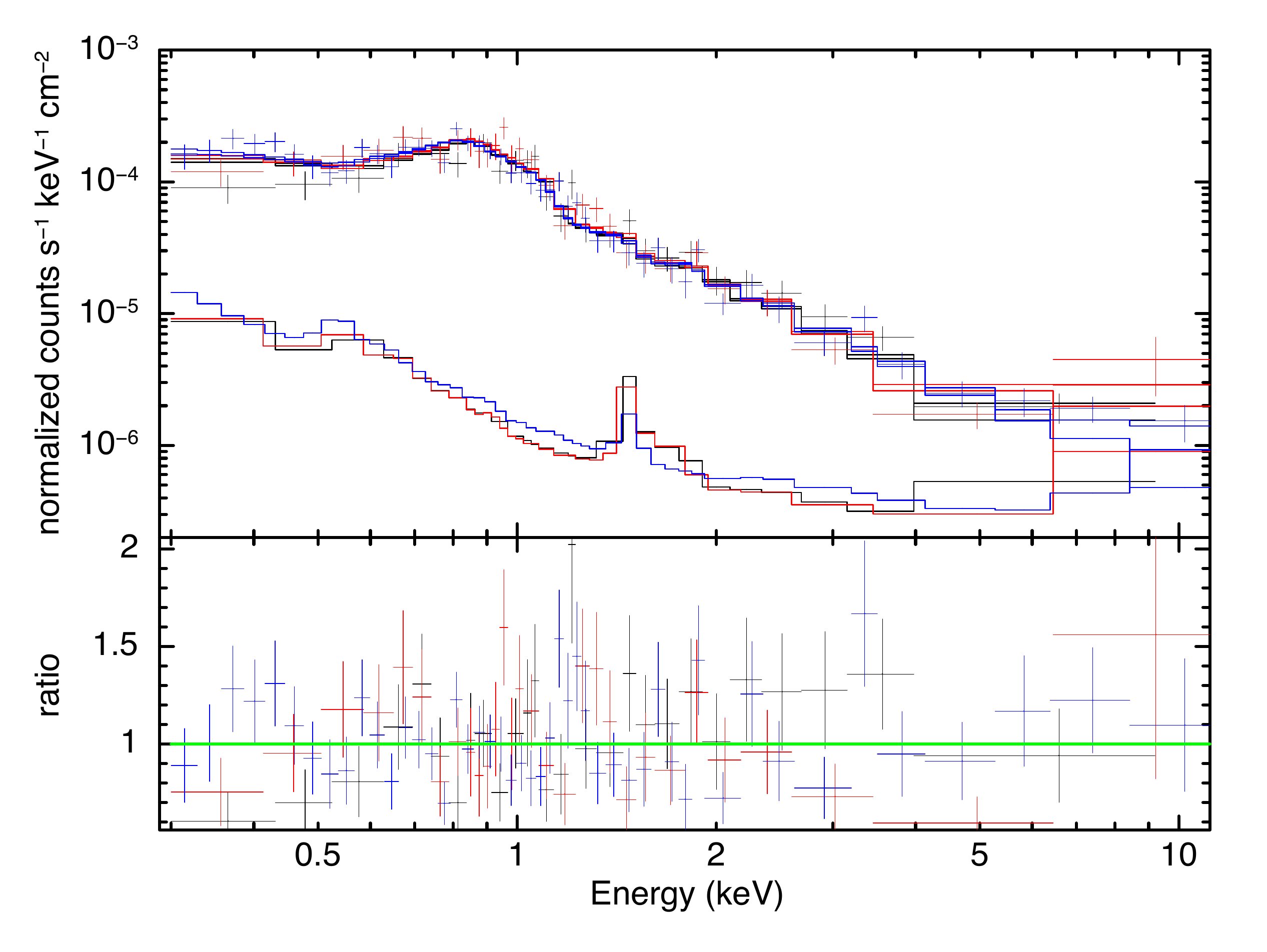}}
\caption{\textit{XMM-Newton} spectral fitting of circular 10 kpc region centered at NGC 5718 with \textit{Model 1} (\textit{top}), \textit{Model 2} (\textit{middle}) and \textit{Model 3} (\textit{bottom}). For plotting purposes only, adjacent bins are grouped until they have a significant detection at least as large as 4$\sigma$, with maximum 10 bins. Upper curve corresponds to the simultaneous fit of the \textit{XMM-Newton} EPIC source and background counts, where the lower curve presents the background model.}
\label{fig:XMMmodels}
\end{figure}

As a first approach, we selected a circular region with \textit{r} = 10 kpc  centered at the BCG with 595 (MOS1), 629 (MOS2), 1481 (PN) net photon counts, which correspond to the 96.3\%, 96.6\%, 96.1\% of total photon counts, respectively. We implemented a single temperature \texttt{apec} model absorbed by a galactic hydrogen column density, that is; \texttt{phabs} $\times$ \texttt{apec} with \textit{XSPEC}, which will be referred to as \textit{Model 1}. This model suggests a \textit{kT} $\simeq$ 1.32 keV plasma where the abundance was fixed to solar abundance 
(\textit{Z$_{\odot}$}). We found C-stat/d.o.f (\textit{C / $\nu$}) of 747.79/189. The simultaneous fit of MOS1, MOS2 and PN spectra is shown in the top panel of Fig.~\ref{fig:XMMmodels}.

Following the spectral fit with a single temperature model, we added a \texttt{powerlaw} component since the spectrum presented in the top panel of Fig.~\ref{fig:XMMmodels} shows the requirement for another component especially beyond 3 keV. The second model, \textit{Model 2}; \texttt{phabs} $\times$ (\texttt{apec} + \texttt{powerlaw}), significantly improved the statistics with \textit{$\Delta$C/$\Delta$$\nu$} $\simeq$ 507.99/2, which can also be seen from the middle panel of Fig.~\ref{fig:XMMmodels}. \textit{Model 2} suggests a \textit{kT} $\simeq$ 0.98 keV plasma in addition to a power law emission with photon index \textit{$\Gamma$} $\simeq$ 1.81, the latter being consistent with an AGN emission. When the abundance was freed to vary, the value was found to be 0.20 $\leqslant$ \textit{Z} $\leqslant$ 1.0 \textit{Z$_{\odot}$} within 90\% confidence levels without affecting the values of the rest of the parameters, therefore was fixed at 1 \textit{Z$_{\odot}$}.

To further improve the fitting of the \textit{XMM-Newton} spectrum, we added a second \texttt{apec} component since this region may enclose a part of the coronal gas and ICM interface region. This additional component further improved the fit, indicating the presence of two thermal plasma components with temperatures \textit{kT} $\simeq$ 0.84 keV and  \textit{kT} $\simeq$ 2.37 keV along with a power-law component with photon index \textit{$\Gamma$} $\simeq$ 1.62, improving the statistics by \textit{$\Delta$C/$\Delta$$\nu$} $\simeq$ 35.21/2. This can be observed from residuals in the energy range between 0.5 - 2 keV in the data to model ratios of the bottom panel of Fig.~\ref{fig:XMMmodels}. This third model; \texttt{phabs} $\times$ (\texttt{apec} + \texttt{apec} + \texttt{powerlaw}), will be referred to as \textit{Model 3}. The spectrum of \textit{Model 3} is shown in the bottom panel of Fig.~\ref{fig:XMMmodels}. 

The best-fit results of these models are summarized in Table~\ref{tab:XMMfit}. We have tested the stability of these models using the spectrum of a slightly smaller region with \textit{r} = 8.5 kpc.

\begin{table}
\caption{Spectral parameters and 1$\sigma$ uncertainty ranges of \textit{XMM-Newton} EPIC spectra for the circular region with \textit{r} = 10 kpc centered at the position of NGC 5718. \texttt{apec} normalization (\textit{norm}) is given in $\frac{10^{-14}}{4\pi \left [ D_A(1+z) \right ]^2}\int n_en_HdV$  where \texttt{powerlaw} normalization ($\kappa$) is \textit{photons} \textit{keV$^{-1}$cm$^{-2}$s$^{-1}$} at 1 keV.}
\label{tab:XMMfit}    
\centering
\begin{tabular}{l r}
\hline\hline
 \\[-0.95em]
 \multicolumn{2}{c}{Model 1: \texttt{phabs} $\times$ \texttt{apec}}\\
 \\[-0.95em]
 \hline
\\[-0.95em]
\textit{kT} (keV) & 1.32 $\pm{0.02}$   \\  
\\[-0.95em]
\textit{Z} (\textit{Z$_{\odot}$}) & 1 (fixed)   \\ 
\\[-0.95em]
\textit{norm} (10$^{-5}$) & 7.67$^{+0.45}_{-0.31}$   \\
\\[-0.95em]
\textit{C / $\nu$} & 747.79/189   \\
\\[-0.95em]
\hline
\\[-0.95em]
\multicolumn{2}{c}{Model 2: \texttt{phabs} $\times$ (\texttt{apec} + \texttt{powerlaw})}\\
\\[-0.95em]
\hline
\\[-0.95em]
\textit{kT} (keV) & 0.98 $\pm{0.03}$   \\  
\\[-0.95em]
\textit{Z} (\textit{Z$_{\odot}$}) & 1 (fixed) \\
\\[-0.95em]
\textit{norm} (10$^{-5}$) & 2.38$^{+0.18}_{-0.14}$   \\
\\[-0.95em]
$\Gamma$  & 1.81$^{+0.05}_{-0.09}$   \\ 
\\[-0.95em]
\textit{$\kappa$} (10$^{-5}$) & 2.28$^{+0.12}_{-0.17}$   \\
\\[-0.95em]
\textit{C / $\nu$} & 239.80/187    \\ 
\\[-0.95em]
\hline
\\[-0.95em]
\multicolumn{2}{c}{Model 3: \texttt{phabs} $\times$ (\texttt{apec} + \texttt{apec} + \texttt{powerlaw})}\\
\\[-0.95em]
\hline
\\[-0.95em]
\textit{kT} (keV) & 0.84$^{+0.08}_{-0.05}$ \\  
\\[-0.95em]
\textit{Z} (\textit{Z$_{\odot}$}) & 1 (fixed) \\
\\[-0.95em]
\textit{norm} (10$^{-5}$) & 1.64$^{+0.12}_{-0.13}$   \\
\\[-0.95em]
$\Gamma$  & 1.62$^{+0.29}_{-0.26}$   \\ 
\\[-0.95em]
\textit{$\kappa$} (10$^{-5}$) & 0.88$^{+0.43}_{-0.34}$   \\
\\[-0.95em]
\textit{kT$_2$} (keV) & 2.37$^{+0.79}_{-0.37}$    \\  
\\[-0.95em]
\textit{$Z_2$} (\textit{Z$_{\odot}$}) & 1 (fixed) \\
\\[-0.95em]
\textit{norm} (10$^{-5}$) & 5.39$^{+1.19}_{-1.43}$   \\
\\[-0.95em]
\textit{C / $\nu$} & 204.59/185    \\
\\[-0.95em]
\hline  
\end{tabular}
\end{table}

\subsection{\textit{Chandra} ACIS-I spectral fitting}
\label{sec:CHANDRAfit}

For the spectral analysis of the \textit{Chandra} observations, we extracted a circular region with \textit{r} = 3 kpc ($\approx$5$\arcsec$) centered at NGC 5718 with 515 net photon counts, which corresponds to the 98.3\% of total photon counts. We used the same application of spectral models as described in Sect.~\ref{sec:XMMfit}; \textit{Model 1}, \textit{Model 2}, \textit{Model 3}. When the abundance value was freed to vary, the value was not constrained and we placed a lower limit of \textit{Z} = 0.3 \textit{Z$_{\odot}$}. The abundance parameters in all models were fixed to 1 \textit{Z$_{\odot}$}, since at \textit{r} = 3 kpc region we expect the coronal gas to have near solar abundance values \citep{Sun07}.

\begin{figure}
\centering
{\includegraphics[width=70mm]{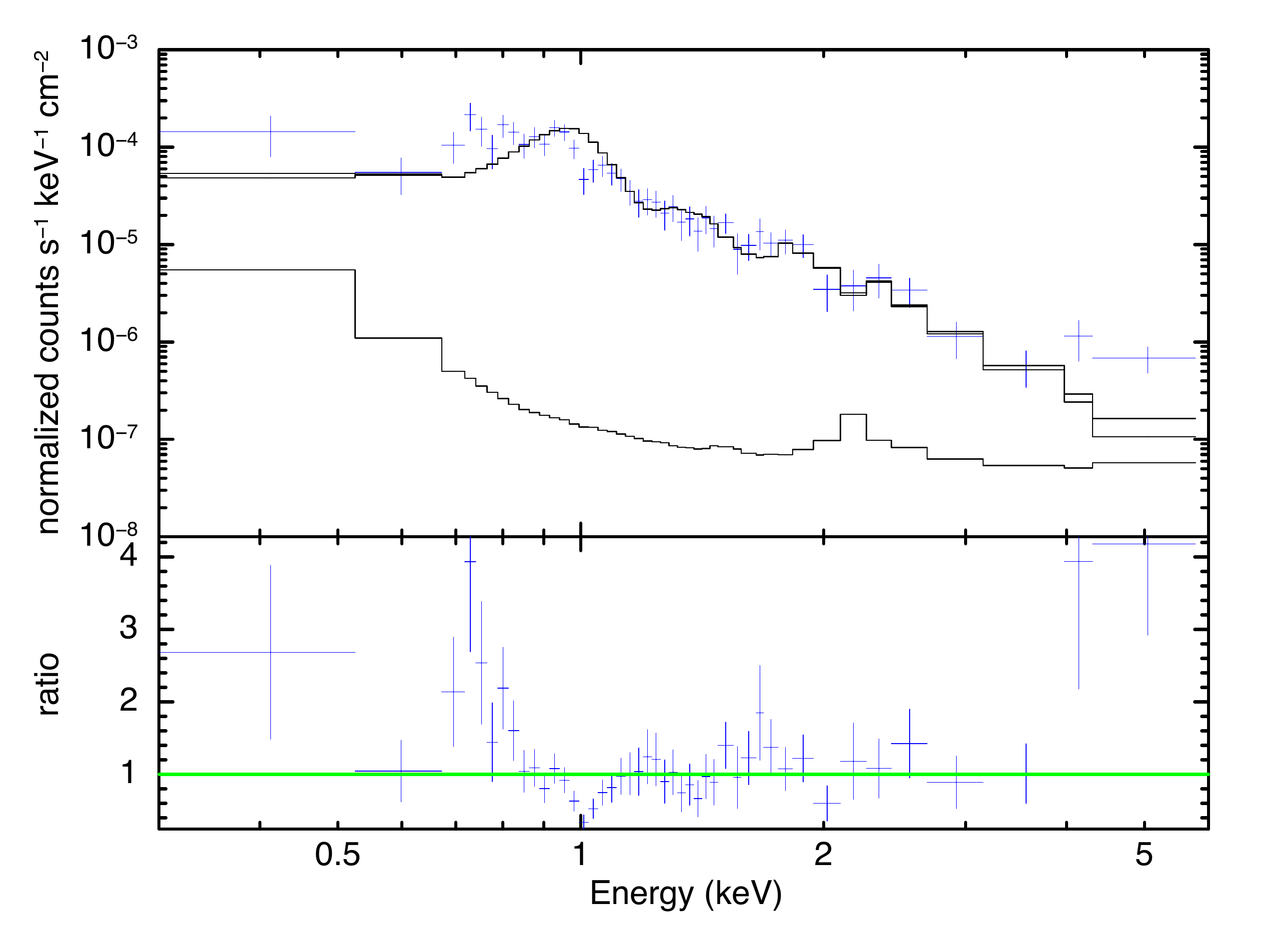}}
{\includegraphics[width=70mm]{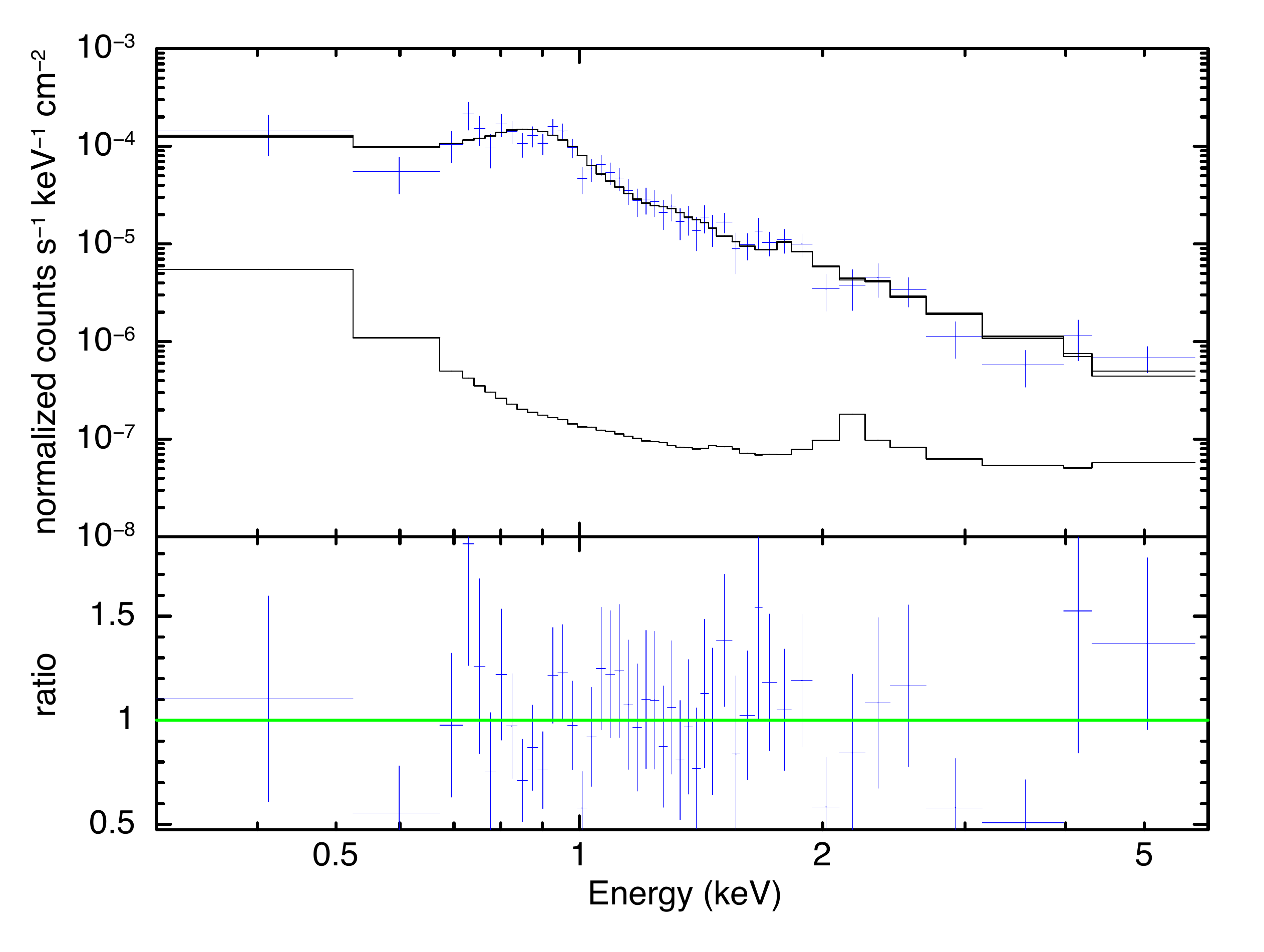}}
\caption{\textit{Chandra} spectral fitting of \textit{r} = 3 kpc circular region centered at NGC 5718 with \textit{Model 1} (\textit{upper}) and \textit{Model 2} (\textit{lower}). For plotting purposes only, adjacent bins are combined until they have a significant detection at least as large as 2$\sigma$, with maximum 5 bins. Upper curve corresponds to the simultaneous fit of the \textit{Chandra} ACIS-I source and background counts, where the lower curve presents the background model.}
\label{fig:Chandramodels}
\end{figure}

As a first step in the spectral analysis, we implemented \textit{Model 1}, that is; \texttt{phabs} $\times$ \texttt{apec}. This model yielded a temperature of \textit{kT} $\simeq$ 1.21 keV, which is slightly lower than the temperature value obtained from \textit{XMM-Newton} using \textit{Model 1}. The corresponding spectrum is presented in the upper panel of Fig.~\ref{fig:Chandramodels}.

The addition of a \texttt{powerlaw} component to \texttt{apec}, \textit{Model 2}, enhanced our statistics by \textit{$\Delta$C/$\Delta$$\nu$} $\simeq$ 72.1/2, which is apparent from the ratio axis shown in the lower panel of Fig.~\ref{fig:Chandramodels}. However, the photon index value of \textit{$\Gamma$} $\simeq$ 2.5 is beyond the value found for \textit{XMM-Newton}. Since the power-law emission from an AGN is more dominant in hard X-ray band, \textit{XMM-Newton} is more sensitive to this emission, hence we implemented a modified \textit{Model 2} for \textit{Chandra} by fixing the value of the photon index to the value obtained from  \textit{XMM-Newton} \textit{Model 2} (Table.~\ref{tab:XMMfit}). Using this method, we found \textit{kT} = 0.93 $\pm{0.04}$, yet \textit{C / $\nu$} = 47.99/43, which is still an improvement of \textit{Model 1} but not of \textit{Chandra} \textit{Model 2}.

The \textit{r} $\simeq$ 10 kpc region selected from \textit{XMM-Newton} required an additional temperature component, which was shown by the improvement of the statistics with an additional \texttt{apec}; namely \textit{Model 3} (Table~\ref{tab:XMMfit}). Unlike the \textit{XMM-Newton} spectral fitting, due to the smaller region selected for the analysis, adding a second \texttt{apec} component to \textit{Chandra} spectrum failed to improve the statistics. In addition, the uncertainty of the temperature values and the photon index increased substantially with the implication of this model. At the radius \textit{r} = 3 kpc, we are well within the NGC 5718. 

The best-fit results of \textit{Model 1} and \textit{Model 2} are summarized in Table~\ref{tab:Chandrafit}. We have tested the stability of these models using the spectrum of a slightly smaller region with \textit{r} = 2.7 kpc.

\begin{table}
\caption{Spectral parameters and 1$\sigma$ uncertainty ranges of \textit{Chandra} ACIS-I spectrum for the circular region with \textit{r} = 3 kpc radius centered at the position of NGC 5718. \texttt{apec} normalization (\textit{norm}) is given in $\frac{10^{-14}}{4\pi \left [ D_A(1+z) \right ]^2}\int n_en_HdV$  where \texttt{powerlaw} normalization ($\kappa$) is \textit{photons} \textit{keV$^{-1}$cm$^{-2}$s$^{-1}$} at 1 keV.}
\label{tab:Chandrafit}    
\centering
\begin{tabular}{l r}
\hline\hline
\\[-0.95em]
 \multicolumn{2}{c}{Model 1: \texttt{phabs} $\times$ \texttt{apec}}\\
 \\[-0.95em]
 \hline
 \\[-0.95em]
\textit{kT} (keV) & 1.21$^{+0.04}_{-0.06}$   \\  
\\[-0.95em]
\textit{Z} (\textit{Z$_{\odot}$}) & 1 (fixed)   \\ 
\\[-0.95em]
\textit{norm} (10$^{-5}$) & 3.25$^{+0.12}_{-0.17}$   \\
\\[-0.95em]
\textit{C / $\nu$} & 111.70/44   \\
\\[-0.95em]
\hline
\\[-0.95em]
\multicolumn{2}{c}{Model 2: \texttt{phabs} $\times$ (\texttt{apec} + \texttt{powerlaw})}\\
\\[-0.95em]
\hline
\\[-0.95em]
\textit{kT} (keV) & 0.91$^{+0.08}_{-0.09}$   \\  
\\[-0.95em]
\textit{Z} (\textit{Z$_{\odot}$}) & 1 (fixed) \\
\\[-0.95em]
\textit{norm} (10$^{-5}$) & 1.51$^{+0.20}_{-0.21}$   \\
\\[-0.95em]
$\Gamma$  & 2.49$^{+0.26}_{-0.34}$  \\ 
\\[-0.95em]
\textit{$\kappa$} (10$^{-5}$) & 1.13$^{+0.25}_{-0.22}$   \\
\\[-0.95em]
\textit{C / $\nu$} & 39.60/42    \\ 
\\[-0.95em]
\hline
\end{tabular}
\end{table}

\subsection{Luminosity}
\label{sec:Luminosity}

In order to assess the contamination of the bremsstrahlung luminosity of the corona - ICM hot gas by the power-law emitting source, we used the \textit{Chandra} observations and analyzed the spectra of both components in concentric annuli of width $\Delta$ r $\sim$ 1 kpc. Firstly, we selected an innermost central circle with \textit{r} $\sim$ 1 kpc and applied \texttt{phabs} $\times$ (\texttt{clumin} $\times$ \texttt{apec} + \texttt{clumin} $\times$ \texttt{powerlaw}) where we set the normalizations of \texttt{apec} and \texttt{powerlaw} to a nonzero value, then \texttt{clumin} normalizations were freed to vary. Then we applied the same model to all successive annuli in order to investigate up to which radius the power-law shows significant contamination in the surface brightness. We calculated the ratio of power-law luminosity with respect to the total luminosity. Our analysis showed that for regions beyond \textit{r} $\simeq$ 3 kpc, the influence of power-law is not significant.

Subsequently, we calculated the luminosity ratio by selecting a new circular region with 0 $\leqslant$ \textit{r} $\leqslant$ 3 kpc. Total luminosity in the energy band 0.5 - 2.5 keV was found to be \textit{L$_{X}$} = 1.07$^{+0.12}_{-0.13}$ $\times$ 10$^{41}$  erg s$^{-1}$, where the power-law component is \textit{L$_{X}$} = 3.04$^{+0.75}_{-0.79}$ $\times$ 10$^{40}$  erg s$^{-1}$. We conclude that power-law accounts for $\sim$30\% of the total luminosity in the central region up to \textit{r} $\simeq$ 3 kpc.

\subsection{Cooling time profile}
\label{sec:cooltime}

We performed an order of magnitude estimation of the cooling time at \textit{r} = 3 kpc centered at NGC 5718 with \textit{Chandra} following the procedure implemented by \citet{Tombesi17}. The normalization of the \textit{XSPEC} model \texttt{apec} is defined as; $\left [10^{-14}/4\pi \left [ D_A(1+z) \right ]^2\right ]\int n_en_HdV$, where the integrant is the emission measure (EM), \textit{n}$_e$ and \textit{n}$_H$ are electron and hydrogen densities in cm$^{-3}$, and the angular diameter distance to NGC 5718 is D$_A$ $\simeq$ 3.55 $\times$ 10$^{26}$ cm. By using the normalization value (1.51$^{+0.20}_{-0.21}$ $\times$ 10$^{-5}$) of the \texttt{apec} model found for \textit{Model 2}, we estimated EM $\simeq$ 2.38$^{+0.31}_{-0.33}$ $\times$ 10$^{63}$ cm$^{-3}$. With the assumption of a fully ionized plasma (\textit{n}$_e$ $\simeq$ 1.2 \textit{n}$_H$) within a spherical volume with \textit{r} $\simeq$ 3 kpc, we estimated \textit{n}$_H$ $\sim$ 0.02 cm$^{-3}$. 

\begin{figure}
\center
{\includegraphics[width=70mm]{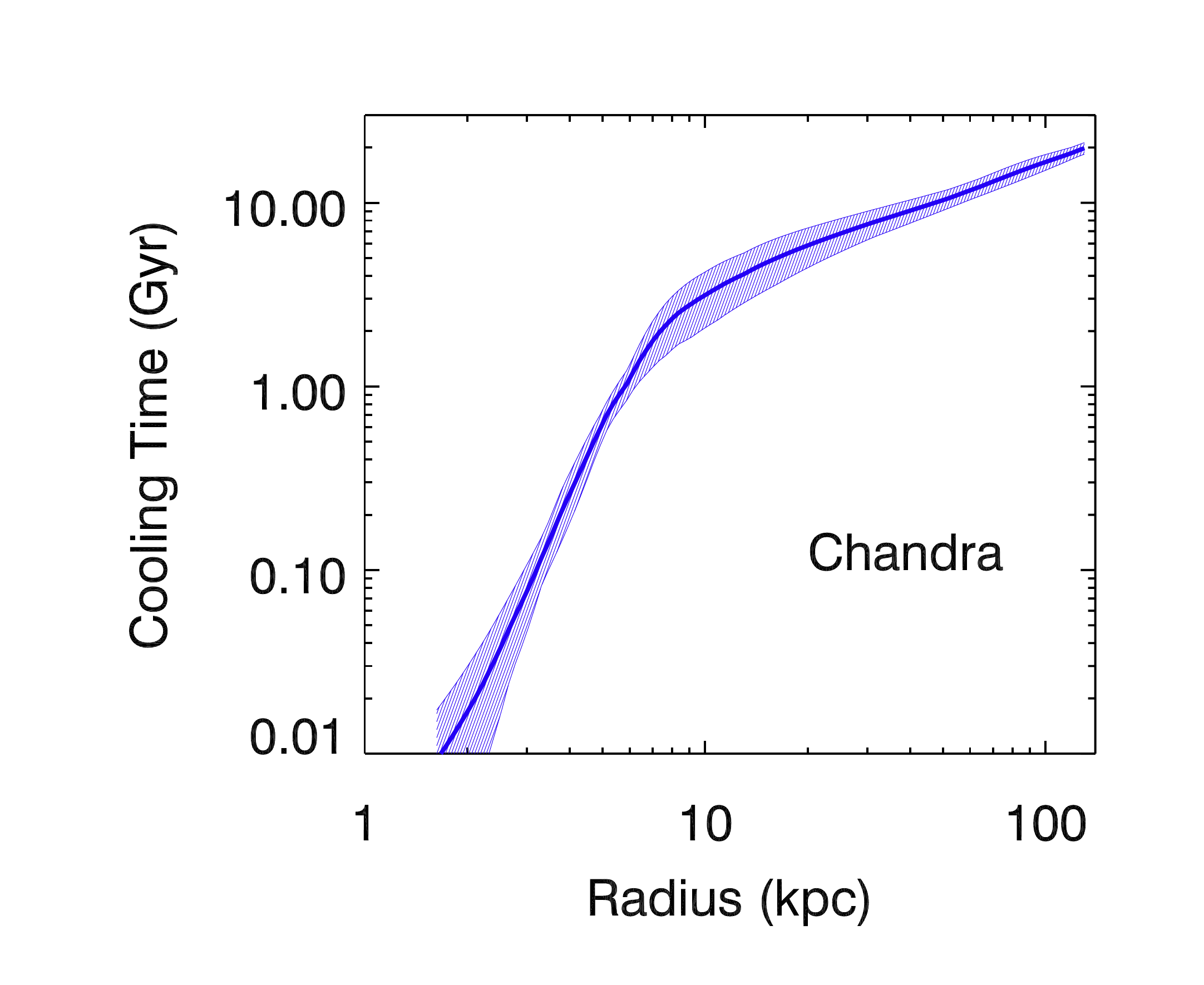}}
\caption{Cooling time profile of MKW 08 obtained from \textit{Chandra} data. Selected region is a disk of \textit{r} = 240$\arcsec$ ($\simeq$ 130 kpc) centered at the BCG.}
\label{fig:tcool}
\end{figure}

Following the method described by \citet{Peterson06}, we estimated the cooling time in the following steps. The X-ray luminosity of the coronal gas within \textit{r} = 3 kpc in the 0.3 - 10 keV energy band of the \texttt{apec} model was found to be \textit{L$_{hot}$} $\simeq$ 1.07 $\times$ 10$^{41}$  erg s$^{-1}$ as described in Sect.~\ref{sec:Luminosity}. Changing the energy band from 0.5 - 2.5 keV to 0.3 - 10 keV did not significantly affect the luminosity. The cooling time of this hot gas of a galaxy can be estimated as \textit{t$_{cool}$} = \textit{U} / \textit{L$_{hot}$} where \textit{U} = (5/2) \textit{NkT} is the internal energy of the gas. Estimating the number of particles, \textit{N}, from the gas density and volume, we calculated \textit{U} $\simeq$ 2.12 $\times$ 10$^{56}$  erg  for \textit{kT} $\simeq$ 1 keV gas at \textit{T} $\simeq$ 10$^{7}$ K. Then, cooling time at \textit{r} = 3 kpc was found to be $\sim$ 64 Myr. By applying this set of calculations using the parameter values obtained from our  density and 3-dimensional temperature profiles, we obtained the cooling time curve for \textit{Chandra}, corresponding to each radial element up to \textit{r} $\simeq$ 130 kpc as shown in Fig.~\ref{fig:tcool}.

\section{Spectroscopic properties of the $\sim$40 kpc tail}
\label{sec:Tail}

\begin{figure}
\center
{\includegraphics[width=70mm]{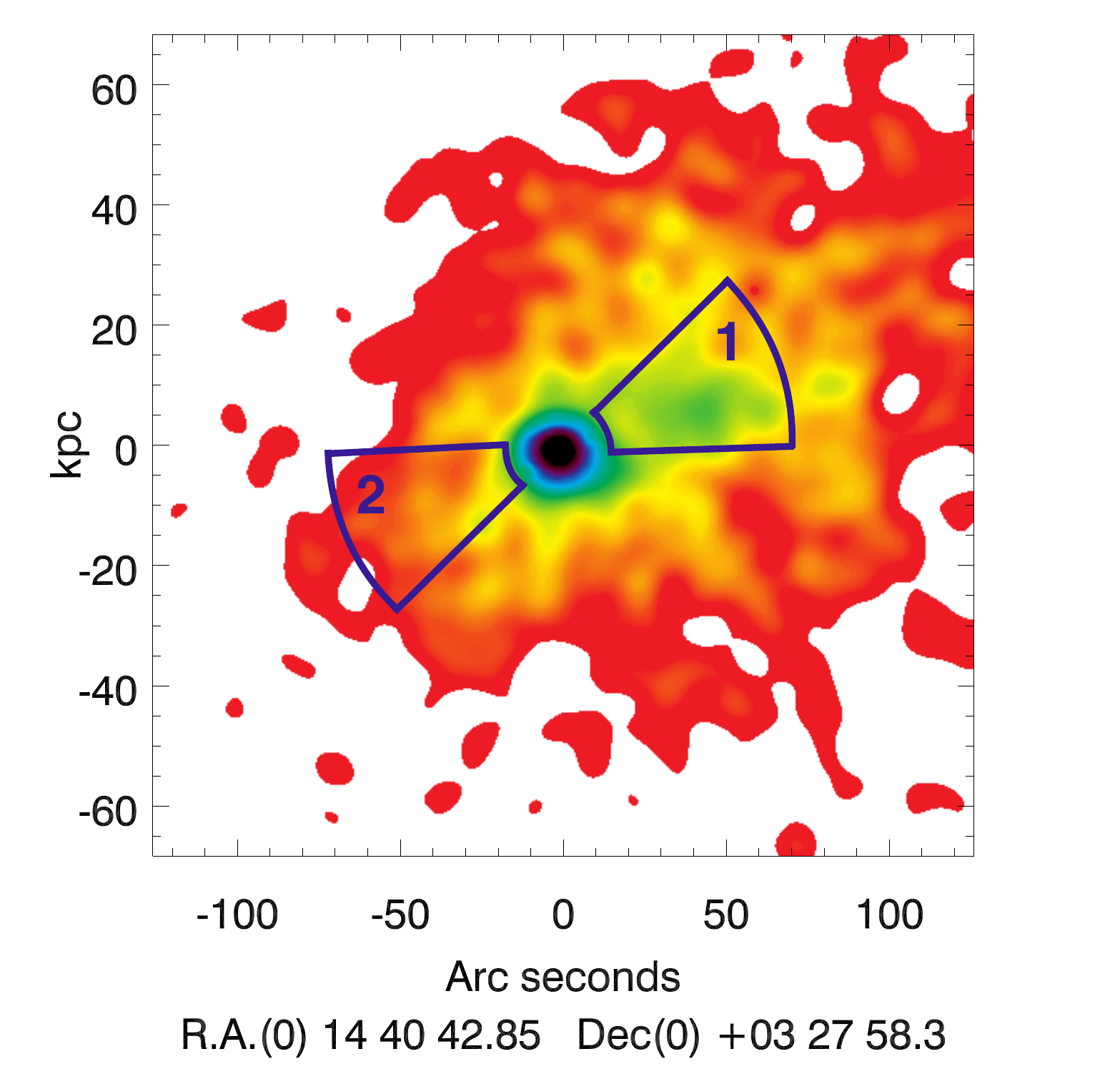}}
\caption{Sectors selected from \textit{Chandra} observations outlined on surface brightness map.}
\label{fig:ChandraSectors}
\end{figure}

The \textit{Chandra} surface brightness map reveals a 40 kpc tail as shown in  the lower right panel of Fig.~\ref{fig:XMMtempmap}. In order to investigate the features of this tail enclosed inside this region, we have selected two sectors from \textit{Chandra}, which are symmetric along the true NE - SW axis as shown in Fig.~\ref{fig:ChandraSectors}. Both sectors cover a 45 degree angle with an inner radius of 10 kpc and an outer radius of 40 kpc. Sector 1 encloses the tail region with 1570 net photon counts corresponding to the 86.2\% of total photon counts, whereas Sector 2 is selected from a region, which does not show contrasting emission features from the surrounding medium, and contains 985 net photon counts corresponding to the 79.8\% of total photon counts. We applied \texttt{phabs} $\times$ \texttt{apec} models to both regions and compared the results. In order to compare the luminosities, we selected 0.5 - 2.5 keV range and applied \texttt{phabs} $\times$ (\texttt{clumin} $\times$ \texttt{apec}) model, where \texttt{apec} normalization was fixed at a nonzero value. We selected the same regions and applied the aforementioned methods to the \textit{XMM-Newton} observation. Sector 1 enclosing the tail region has 547 (MOS1), 468 (MOS2), 1278 (PN) net photon counts, which correspond to the 92.6\%, 91.9\%, 91.8\% of total photon counts, respectively. Whereas, Sector 2 has 357 (MOS1), 333 (MOS2), 820 (PN) net photon counts, which correspond to the 88.8\%, 88.3\%, 88.4\% of total photon counts, respectively.
Results are presented in Table~\ref{tab:Sectorfit} and the corresponding spectra are given in Fig.~\ref{fig:Sectors} and Fig.~\ref{fig:SectorsXMM} in the appendix section of our work.

Although temperature values of Sector 1 and Sector 2 agree within 2$\sigma$ errors, both the spectral fit and parameter values show that the NGC 5718 tail (Sector 1) encloses a medium with high luminosity and abundance with respect to Sector 2, while the \textit{XMM-Newton} spectral analysis results provide a better constrained abundance values.

\begin{table}
\caption{Spectral parameters of \textit{Chandra} ACIS-I spectrum for the selected sectors shown in Fig.~\ref{fig:ChandraSectors}. Errors are presented within 1$\sigma$. \texttt{apec} normalization (\textit{norm}) is given in $\frac{10^{-14}}{4\pi \left [ D_A(1+z) \right ]^2}\int n_en_HdV$. Luminosity values were calculated within 0.5 - 2.5 keV energy band.}
\label{tab:Sectorfit}    
\centering
\begin{tabular}{l c c}
\hline\hline
 \\[-0.95em]
 \textit{Sector 1} & \textit{Chandra} & \textit{XMM-Newton} \\  
 \\[-0.95em]
 \hline
 \\[-0.95em]
\textit{kT} (keV) & 4.65$^{+0.39}_{-0.36}$  & 3.62 $\pm{0.23}$\\  
\\[-0.95em]
\textit{Z} (\textit{Z$_{\odot}$}) & 0.90$^{+0.35}_{-0.22}$ & 0.49$^{+0.13}_{-0.11}$  \\
\\[-0.95em]
\textit{norm} (10$^{-4}$) & 1.30$^{+0.08}_{-0.11}$  & 1.48$\pm{0.08}$ \\
\\[-0.95em]
\textit{C / $\nu$} & 61.69/87  &  205.18/198  \\
\\[-0.95em]
\textit{Luminosity} (10$^{41}$ erg s$^{-1}$) & 1.68$\pm{0.05}$ & 1.76$\pm{0.08}$ \\
\hline
\\[-0.95em]
 \textit{Sector 2} & \textit{Chandra} & \textit{XMM-Newton}\\  
\\[-0.95em]
\hline
\\[-0.95em]
\textit{kT} (keV) & 4.09$^{+0.48}_{-0.43}$ & 2.90$^{+0.26}_{-0.28}$ \\  
\\[-0.95em]
\textit{Z} (\textit{Z$_{\odot}$}) & 0.22$^{+0.19}_{-0.16}$ & 0.20$^{+0.10}_{-0.09}$\\
\\[-0.95em]
\textit{norm} (10$^{-4}$) & 1.06$^{+0.07}_{-0.08}$  & 0.90$\pm{0.07}$\\
\\[-0.95em]
\textit{C / $\nu$} & 39.60/42   &  155.45/163 \\ 
\\[-0.95em]
\textit{Luminosity} (10$^{41}$ erg s$^{-1}$) & 1.16$\pm{0.05}$ & 0.96$\pm{0.06}$\\
\hline
\end{tabular}
\end{table}

\section{Discussion}

The temperature and pressure profiles obtained from the \textit{Chandra} data reveal three characteristic regions of the corona (\textit{r} $\leqslant$ 4 kpc), the corona - ICM interface (4 $\leqslant$ \textit{r} $\leqslant$ 10 kpc) and the ICM (10 kpc $\leqslant$ \textit{r}). The density and temperature profiles, as well as the characteristic spatial scale (\textit{r} $\leqslant$ 4 kpc) of the corona are typical of luminous X-ray thermal coronae of galaxies in hot clusters, as reported for seven of the most luminous coronae in the systematic study of 25 hot nearby clusters of \citet{Sun07}. NGC 5718 is classified as a BCG corona in literature \citep{Sun07} whose thermal emission is found to be of ISM origin. Our results further support their findings that the hot ISM in BCG coronae indeed survive ICM stripping since otherwise we would expect smooth thermodynamical profiles at cluster scales due to pure cooling of ICM \citep[see, e.g.,][]{Gaspari12} rather than sharp edges at \textit{r} $\sim$ 4 kpc, then again at \textit{r} $\sim$ 10 kpc. In addition, the \textit{Chandra} temperature profile hints at a temperature gradient inside 4 $\leqslant$ \textit{r} $\leqslant$ 10 kpc interface, and the high pressure ISM seems to reach pressure equilibrium with the surrounding ICM inside this interface. 

In both spectral analyses of the \textit{XMM-Newton} and \textit{Chandra} data, a power-law emission was required besides the hot gas emission, for the region centered at NGC 5718.  The power-law has photon indices of \textit{$\Gamma$} = 1.81$^{+0.05}_{-0.09}$ and \textit{$\Gamma$} =  2.49$^{+0.26}_{-0.34}$ for \textit{XMM-Newton} and \textit{Chandra} respectively, which are consistent within the 2$\sigma$ confidence range. However, due to superior energy sensitivity of \textit{XMM-Newton} with respect to \textit{Chandra} at energies higher than 5 keV, and photon index precision, it is more likely that the power-law emission of \textit{XMM-Newton} is more accurate, as demonstrated also by the much smaller error bars. This finding suggests the existence of an AGN along with the findings of \citet{Hogan15} who find evidence for radio activity of this BCG, and of \citet{Bharadwaj14} who point to the central radio source in MKW 08. The photon index of 1.8 $\leqslant$ \textit{$\Gamma$} favor an AGN emission rather than X-ray binaries (\textit{$\Gamma$} $\leqslant$ 1.4) \citep[see, e.g.,][]{Tozzi06}. In addition, the difference between the photon indices obtained from \textit{XMM-Newton} and \textit{Chandra} may be due to the intrinsic temporal variability of the AGN \citep[see, e.g.,][]{Fabian15}, as the observation date of the \textit{XMM-Newton} and \textit{Chandra} data span more than 10 years. We investigated this feature by inserting the \textit{XMM-Newton} photon index value into the \textit{Model 2} of merged \textit{Chandra} observations as described in Sect.~\ref{sec:CHANDRAfit} and found that BCG ISM gas is not affected by this intrinsic temporal variability.
    
Thanks to the \textit{Chandra} spectral analysis, for the circular region of \textit{r} $\simeq$ 3 kpc centered at the BCG, we found a temperature value of \textit{kT} $\simeq$ 1 keV, which agrees with a hot ISM emission. An additional thermal component did not improve the statistics, indicating that in the central $\simeq$ 3 kpc region, the thermal emission is not substantially influenced by the emission from the interface region between the corona (a hot ISM or a mini-cool core) of NGC 5718 and the ICM. For the \textit{r} $\simeq$ 3 kpc region, we found a cooling time value of $\sim$ 60 Myr, which shows that although classified as a NCC by its central cooling time at \textit{r} = 0.4\% R$_{500}$ \citep{Hudson10}, MKW 08 hosts a mini cool core at a much smaller spatial scale (\textit{r} $\simeq$ 3 kpc) as opposed to CC clusters, which agrees with the findings of \citet{Sun07}. The cooling time value of $\sim$ 60 Myr is three orders of magnitude smaller than the Hubble time, which implies that there is a need for a heating mechanism to keep the ISM hot, otherwise the gas enclosed in this region would have experienced a pure cooling flow catastrophe, leading to the collapse of the inner gaseous atmosphere. 

Star formation, AGN heating, and heat conduction from the ICM are possible heating mechanisms, which can hinder the cooling of the hot ISM. In order to investigate the most possible scenario of the heating mechanism, firstly we estimated the star formation rate expected to sustain the heating of the X-ray emitting gas of NGC 5718. Star formation is expected to heat the gas mostly through shocks from supernova and stellar winds then the gas cools down via bremsstrahlung \citep[see, e.g.,][]{Persic02}. Using the relation of X-ray luminosity and star formation rate given by \citet{Mineo14}, \textit{L$_{X(0.5-8.0 keV)}$} $\simeq$ 4 $\times$ 10$^{39}$ ($\dot{M}_\ast / \textit{M$_\odot$} yr^{-1}$), we found a star formation rate of $\dot{M}_\ast$ $\sim$ 40 \textit{M$_{\odot}$} yr$^{-1}$. This value is more than an order of magnitude larger than the star formation rate expected from radio galaxies \citep[see, e.g.,][]{Tombesi17,Westhues16}, and is not likely adequate to supply enough heating to balance the cooling. 

Considering also the deprojected temperature profiles shown in Fig.~\ref{fig:ktprof} where the hot ISM gas is isothermal, we conclude that heat conduction does not play a role as a heating mechanism, and AGN is the most likely heating mechanism. Indeed, the presence of a mini cool core (even in a macroscale NCC) may imply that in the past and/or near future, a rain of cold gas has been or will condense out of the plasma atmosphere, feeding the SMBH via chaotic cold accretion, hence boosting its accretion rate and thus the injection of AGN outflows or jets \citep{Gaspari18}. This induces a duty cycle with frequency tied to the central cooling time.

A hot ISM with a temperature of \textit{kT} $\sim$ 1 keV located within $\sim$ 4 kpc of the AGN is recently reported also in the \textit{Chandra} grating spectra of two bright radio galaxies, namely 3C 390.3 and 3C 120 \citep{Tombesi16b,Tombesi17}. These radio galaxies show evidence of both powerful wind and jets, and it is suggested that their mechanical power may be enough to provide a heating source for the hot ISM \citep[see, e.g.,][]{Tombesi17}. Theories of AGN feedback driven by winds or jets predicts the existence of hot shocked bubbles \citep[see, e.g.,][]{Wagner12,Wagner13,Zubovas14}. The forward shock decelerates while the AGN wind sweeps up the ambient medium, resulting in a cooler gas with a temperature of T $\sim$ 10$^{7}$ K for a shock velocity of $\sim$ 1000 km s$^{-1}$. Depending on the power of the central AGN, the X-ray luminosity of the thermal bremsstrahlung expected from wind-shocked gas is \textit{L$_{X}$} $\sim$ 10$^{41}$-10$^{42}$ erg s$^{-1}$ \citep{Bourne13, Nims15}. These parameters are overall consistent with estimates for the hot X-ray-emitting gas in the central 3 kpc region of NGC 5718. 

Luminosity deduced from the \textit{Chandra} data inside the central \textit{r} $\simeq$ 3 kpc region in the 0.5 - 8.0 keV band was found to be \textit{L$_{X}$} $\simeq$ 1.32 $\times$ 10$^{41}$ erg s$^{-1}$. We can exclude an X-ray emission from an ultraluminous X-ray source (ULX) within 3 kpc, due to the fact that the emission is consistent with being centered at the galaxy NGC 5718 and the X-ray luminosity of $\sim$ 1.32 $\times$ 10$^{41}$ erg s$^{-1}$ would imply a too high black hole mass of >1000 \textit{M$_{\odot}$} if accreting at the Eddington level or a stellar mass black hole accreting at an extreme level, >1000 the Eddington limit \citep[see, e.g.,][]{Kaaret17}. 

Using our estimated density value in Sect. \ref{sec:cooltime}, as well as adopting the gas mass (\textit{M$_{gas}$}) - density equation by \citet{Ettori09}, we estimated the \textit{M$_{gas}$} of the coronal gas inside \textit{r} $\simeq$ 3 kpc with the assumption of a constant density within a spherical volume. The resulting value, \textit{M$_{gas}$} $\sim$ 6.17 $\times$ 10$^{7}$ \textit{M$_{\odot}$}, is in good agreement with the average gas mass estimates for a sample of X-ray coronae as reported by \citet{Sun07}, as well as with the two BCGs of the Coma cluster \citep{Vik01}.

In a systematic analysis of embedded X-ray coronae inside 25 hot clusters, \citet{Sun07} reported a fraction of X-ray tails as small as 5\% of their population of 76 early-type galaxy coronae. If this is indeed the case, such a scarcity of X-ray tails could imply quick ISM-ICM mixing mechanisms, or rare stripping phases. Considering only one short \textit{Chandra} observation of MKW08, they did not report a tail linked to the BCG corona. However, with deeper \textit{Chandra} observations, we could show that NGC 5718 indeed has a $\sim$ 40 kpc tail. Our analysis shows that the region enclosing this tail described as Sector 1 shown in Fig.~\ref{fig:ChandraSectors}, has a high abundance value (\textit{Z} $\simeq$ 0.9 \textit{Z$_{\odot}$} provided by the \textit{Chandra} spectrum and \textit{Z} $\simeq$ 0.5 \textit{Z$_{\odot}$} from the \textit{XMM-Newton} spectrum) compared to Sector 2 (\textit{Z} $\simeq$ 0.2 \textit{Z$_{\odot}$}). This abundance value is consistent with a typical hot ISM abundance, which suggests that this tail has been formed by the stripped material from the BCG. The morphological coincidence of this tail with a bridge connecting NGC 5718 and IC 1042 in the optical image of Fig.~\ref{fig:optical} let us speculate that an interaction process between NGC 5718 and IC 1042 is partly responsible for this stripping. We analyzed the \textit{XMM-Newton} and \textit{Chandra} photon images using various kind of smoothing thresholds. It appeared that the BCG tail is very faint and overlaps with a more extended elongation that is best evidenced with the \textit{XMM-Newton} imaging analysis. It is likely that the \textit{XMM-Newton} elongation, which shows a different direction from the \textit{Chandra} images, reflects the cluster morphology independently from the BCG tail. In contrast with what is observed toward NGC 5718, the lack of any visible X-ray emission toward IC 1042 is an open issue. It is interesting to remember in this respect that only 60\% of luminous galaxies in the sample of \citet{Sun07} were reported to host X-ray coronae.

\section{Conclusion}

In this work we present a spectroscopic and imaging study of the galaxy cluster MKW 08 from the central BCG, NGC 5718, to the ICM up to $\simeq$ 420 kpc using archival \textit{XMM-Newton} EPIC and \textit{Chandra} ACIS-I observations. We found that NGC 5718 hosts a hot ISM (\textit{kT} $\simeq$ 1 keV) and a central AGN with photon index  \textit{$\Gamma$} $\simeq$ 1.8. We located a BCG corona (a hot ISM or a mini cool core) and ICM interface in the central 4 $\leqslant$ \textit{r} $\leqslant$ 10 kpc region. Furthermore, we argued that the possible heating mechanism for fueling the hot ISM of the BCG corona is the AGN heating. Thanks to the $\sim$ 100 ks \textit{Chandra} exposure, we also discovered that the BCG has formed a chemically rich $\sim$ 40 kpc tail. In conclusion, this study suggests that high spatial and spectral resolution is required to improve the understanding of the evolution of BCGs and their interaction with their host clusters. Hitomi results on Perseus \citep{Hitomi16} indicate that the spectral-imaging method employed in this work will have implications to study the interrelation of BCG corona and ICM in detail with future missions such as \textit{XRISM} \citep{xrism}, \textit{Lynx} \citep{lynx}, \textit{AXIS} \citep{axis} and \textit{Athena} \citep{athena}.

\begin{acknowledgements} 
The authors thank the anonymous referee for suggestions that led to important improvements in the paper. The research leading to these results has received funding from the European Union's Horizon 2020 Programme under the AHEAD project (grant agreement n. 654215). A.T. and E.N.E. would like to thank Bo\u{g}azi\c{c}i University Research Fund Grant Number 13760 for financial support. A.T. acknowledges support by the Study in Italy Grant, Research Under Academic Supervision. F.T. acknowledges support by the Programma per Giovani Ricercatori - anno 2014 "Rita Levi Montalcini". M.G. is supported by the Lyman Spitzer Jr. Fellowship (Princeton University) and by NASA Chandra GO7-18121X and GO8-19104X. R.S. acknowledges financial contribution from the agreement ASI-INAF n.2017-14-H.0. The authors thank Massimo Cappi for useful discussions. A.T. thanks Fran\c{c}ois Mernier and Jelle de Plaa for useful discussions.
\end{acknowledgements}

\bibliographystyle{aa}
\bibliography{MKW08}

\begin{appendix}
\section{The double $\beta$-model: parameters and fit results}

In this section, we present the simultaneous fit results of the surface brightness and temperature profiles of \textit{XMM-Newton} and \textit{Chandra} obtained by using Eqn.~\ref{eqn:3dGasDensity} and Eqn.~\ref{temp}. The reader should consider the resulting values only as a representative since the double $\beta$-model comprising the description of both the corona and the ICM as well as a correction for the central power cusp feature deviates from the original $\beta$-model proposed by \citet{Cavaliere}, that is to say; the $\beta$ parameter is designated to the slope of the electron density profile in our treatment instead of bearing its original description {$\beta = \mu m_{p} \sigma^2 / kT_{gas}$} referring to isothermal spheres. In essence, the purpose of this treatment is to find a smooth analytical function, which describes the electron density profile and also conserves its shape when integrated. The parameters are degenerate, meaning different values for several parameters can result in the same curve, which does not affect the integration along the line of sight, and the given $\beta$ values by themselves should not be used to estimate physical properties.

The definitions of the parameters of Eqn.~\ref{eqn:3dGasDensity} are given as;
 \[
\begin{array}{lp{0.8\linewidth}}
 n_{e}  & electron density     \\
 n_{p}               & proton density                   \\
n_{0}            & central number density of the ICM component            \\
n_{02}         & central number density of the corona component            \\
r          & radius (independent variable)   \\
r_{c}             & core radius of the ICM distribution \\
r_{c2}             & core radius of the corona distribution\\
\alpha          & central power-law cusp index            \\
\gamma          & width of the transition region              \\
\varepsilon         & slope change near the radius $r_{s}$            \\
\beta          & slope of the electron density profile               \\
 \end{array}
\]

\begin{table}[h!]
\caption{Best-fit results of \textit{XMM-Newton} and \textit{Chandra} corresponding to Fig.~\ref{fig:XMMbrightness}, Fig.~\ref{fig:Chandrabrightness} and Fig.~\ref{fig:ktprof}}
\label{tab:allfit}    
\centering
\begin{tabular}{l c c}
\hline\hline
 & \textit{XMM-Newton} &  \textit{Chandra} \\  
 \hline
  \\[-0.8em]
\textit{n$_{0}$}  (10$^{-3}$ cm$^{-3}$)  &2.37$^{+0.00745}_{-0.0859}$    &    17.6$^{+0.218}_{-0.110}$    \\
 \\[-0.8em]
n$_{02}$ (cm$^{-3}$)      & 0.198$^{+0.0358}_{-0.0317}$ & 7.41$^{+0.885}_{-0.853}$       \\
 \\[-0.8em]
r$_c$     (10$^{-1}$ kpc)         & 434$^{+42.8}_{-7.42}$ &  12.9$^{+0.296}_{-0.0884}$ \\
 \\[-0.8em]
r$_{c2}$  (10$^{-1}$ kpc)         & 12.3$^{+5.09}_{-0.693}$ & 9.64$^{+0.373}_{-0.226}$\\
 \\[-0.8em]
r$_{s}$ (Mpc)             & 0.997$^{+0.0509}_{-0.0244}$ & 1.51$^{+0.158}_{-0.0916}$ \\
 \\[-0.8em]
$\alpha$          &  0.628$^{+0.104}_{-0.115}$ &     0$^{+1.32}_{-0}$      \\
 \\[-0.8em]
$\beta$   (10$^{-2}$)      & 20.0$^{+0}_{-0}$  &  20.0$^{+0.219}_{-0}$  \\
 \\[-0.8em]
$\beta$$_{2}$  (10$^{-1}$)         & 5.45$^{+0.386}_{-0.191}$ &  12.5$^{+0.396}_{-0.371}$ \\
 \\[-0.8em]
$\gamma$   (10$^{-1}$)        & 15.7$^{+0.356}_{-0.454}$ &    12.1$^{+0.624}_{-0.729}$     \\
 \\[-0.8em]
$\varepsilon$          & 13.2$^{+0.387}_{-0.0272}$ &   8.52$^{+0.479}_{-0.641}$    \\
 \\[-0.8em]
T$_{0}$ (keV)            & 3.73$^{+0.891}_{-0.320}$ & 4.01$^{+ 0.364}_{-0.0920}$ \\
 \\[-0.8em]
r$_{t}$     (10$^{-2}$ Mpc)      & 12.6$^{+25.2}_{-1.99}$ &5.7$5^{+2.68}_{-0.196}$\\
 \\[-0.8em]
a     (10$^{-1}$)     &  -1.04$^{+0.761}_{-0.516}$ & 0.421$^{+0.388}_{-0.679}$\\
 \\[-0.8em]
b             &2.67$^{+2.39}_{-0.605}$ & 0.100$^{+2.13}_{-0}$\\
 \\[-0.8em]
c            &0.573$^{+1.29}_{-0.143}$ & 0$^{+0.119}_{-0}$\\
 \\[-0.8em]
 \hline
\end{tabular}
\end{table}

\section{Count rates}

In this section, we provide the count rates for the temperature bins corresponding to Fig.~\ref{fig:ktprof}. The temperature profiles were constructed by using eleven bins for the \textit{XMM-Newton} EPIC and ten bins for the \textit{Chandra} ACIS-I cameras.

\begin{table*}
\caption{Estimated photon counts of \textit{XMM-Newton} and \textit{Chandra} within 0.3-12.0 keV band for the bins shown in Fig.~\ref{fig:ktprof}. Percentages are the net photon counts with respect to the total photon counts including background photons.}
\label{tab:Countrates}    
\centering
\begin{tabular}{lrrrr}
\hline\hline
\\[-0.95em]
Regions &  \multicolumn{4}{c}{Net photon counts}\\
  \hline
 &EPIC-MOS1&EPIC-MOS2&EPIC-PN & ACIS-I\\
 \hline 
Region 1 & 113 (98.3\%)&129 (98.5\%)& 240 (98.4\%)& 162 (99.4\%)\\ 
Region 2 & 150 (97.4\%)& 162 (97.6\%)& 371 (97.4\%)& 101 (99.0\%)\\ 
Region 3 & 292 (95.5\%)&215 (96.0\%)& 542 (95.8\%)& 131 (97.8\%)\\ 
Region 4 & 323 (93.1\%)&339 (93.4\%)& 847 (93.1\%)& 110 (96.5\%)\\ 
Region 5 & 729 (91.8\%)& 728 (91.9\%)& 1773 (91.3\%)& 300 (92.6\%)\\ 
Region 6 & 1572 (89.9\%)& 1414 (89.4\%)& 3444 (89.1\%)& 601 (87.4\%)\\ 
Region 7 & 3125 (87.0\%)&3121 (87.2\%)&6455 (86.7\%)& 1866 (84.6\%)\\ 
Region 8 & 5734 (82.0\%)&5835 (82.6\%)& 12961 (82.4\%)& 5362 (81.1\%)\\ 
Region 9 &8782 (73.9\%)&9619 (75.4\%)&19021 (73.6\%)& 13134 (74.8\%)\\ 
Region 10 & 7876 (58.7\%)&9320 (57.4\%)& 22594 (56.3\%)& 29554 (65.2\%)\\ 
Region 11 & 4651 (28.0\%) &5603 (26.9\%)& 14711 (27.4\%)&  - \\ 
\hline
\end{tabular}
\end{table*}

\section{Spectra of the $\sim$40 kpc tail}

The resulting spectra from the sectors of \textit{XMM-Newton} and \textit{Chandra}, which are analyzed in Sect.~\ref{sec:Tail}, are presented in this section. Sector 1 is the region, which encloses the tail region, whereas Sector 2 was selected from a symmetrical region along the true northeast - southwest axis as indicated in Fig.~\ref{fig:ChandraSectors}.

\begin{figure}
\centering
{\includegraphics[width=70mm]{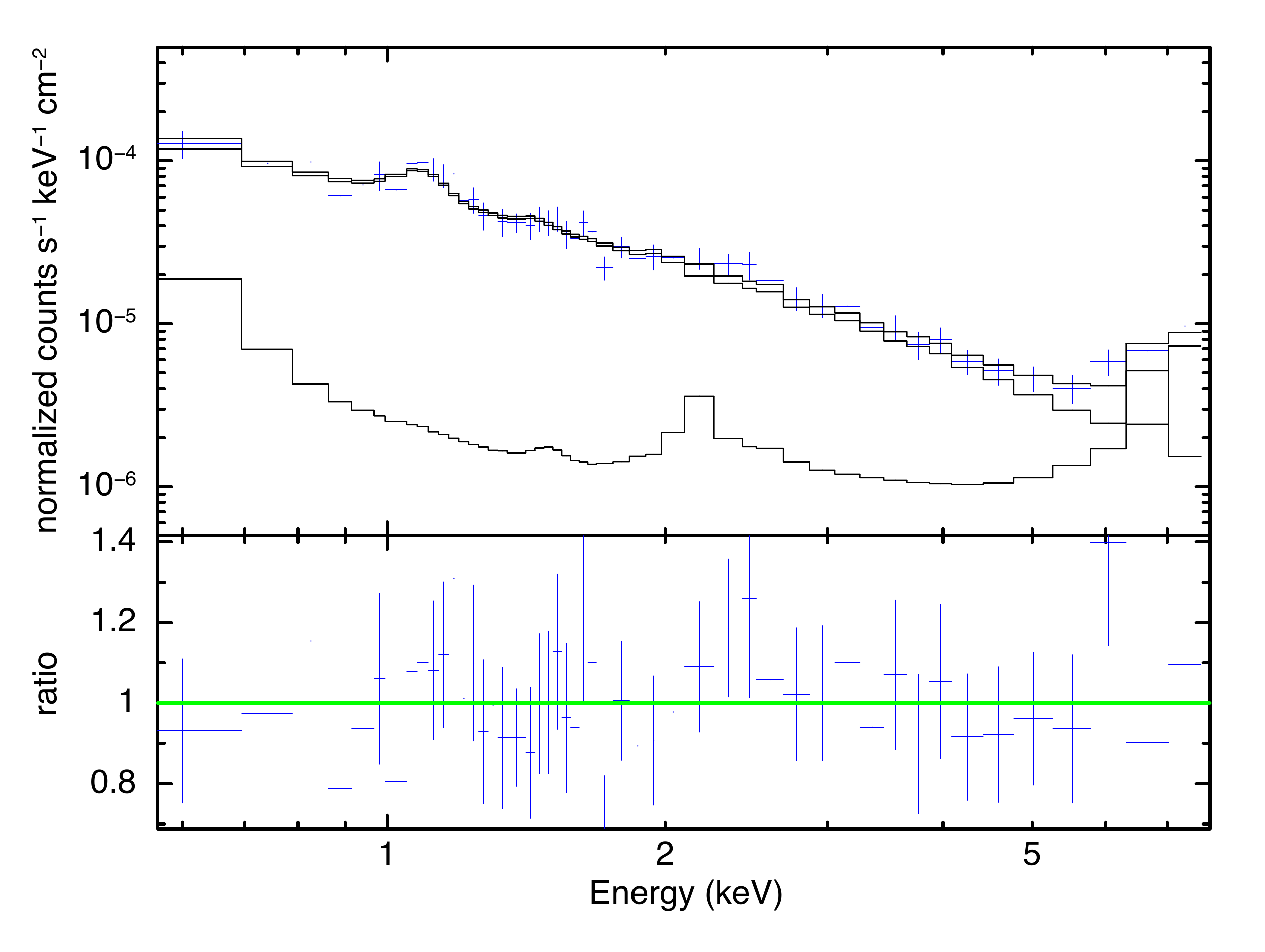}}
{\includegraphics[width=70mm]{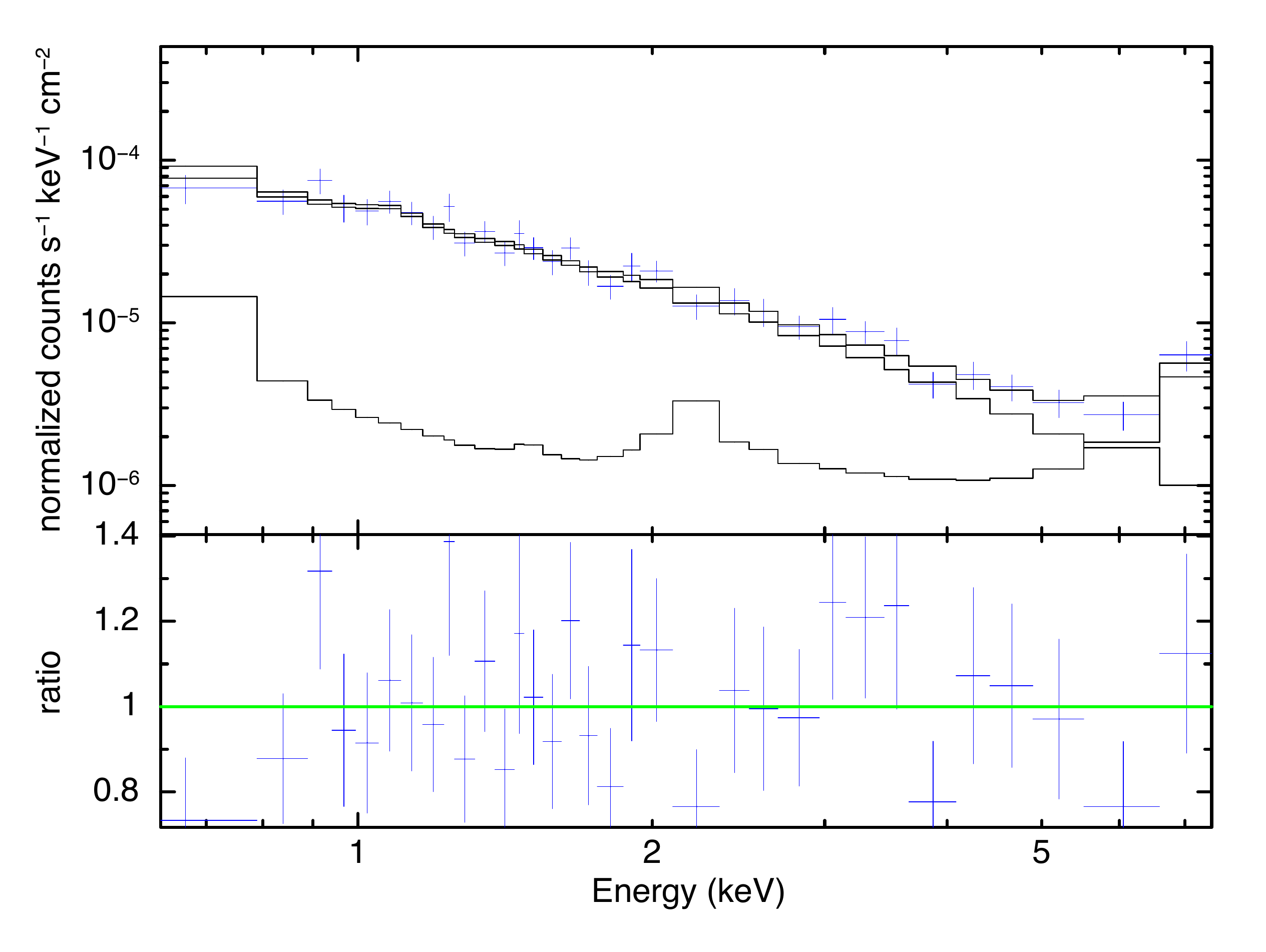}}
\caption{\textit{Chandra} spectral fitting of Sector 1 (\textit{upper}) and Sector 2 (\textit{lower}) with \texttt{phabs} $\times$ \texttt{apec} models. For plotting purposes only, adjacent bins are combined until they have a significant detection at least as large as 5$\sigma$, with maximum 10 bins. Upper curve corresponds to the simultaneous fit of the \textit{Chandra} ACIS-I source and background counts, where the lower curve presents the background model in each panel.}
\label{fig:Sectors}
\end{figure}

\begin{figure}
\centering
{\includegraphics[width=70mm]{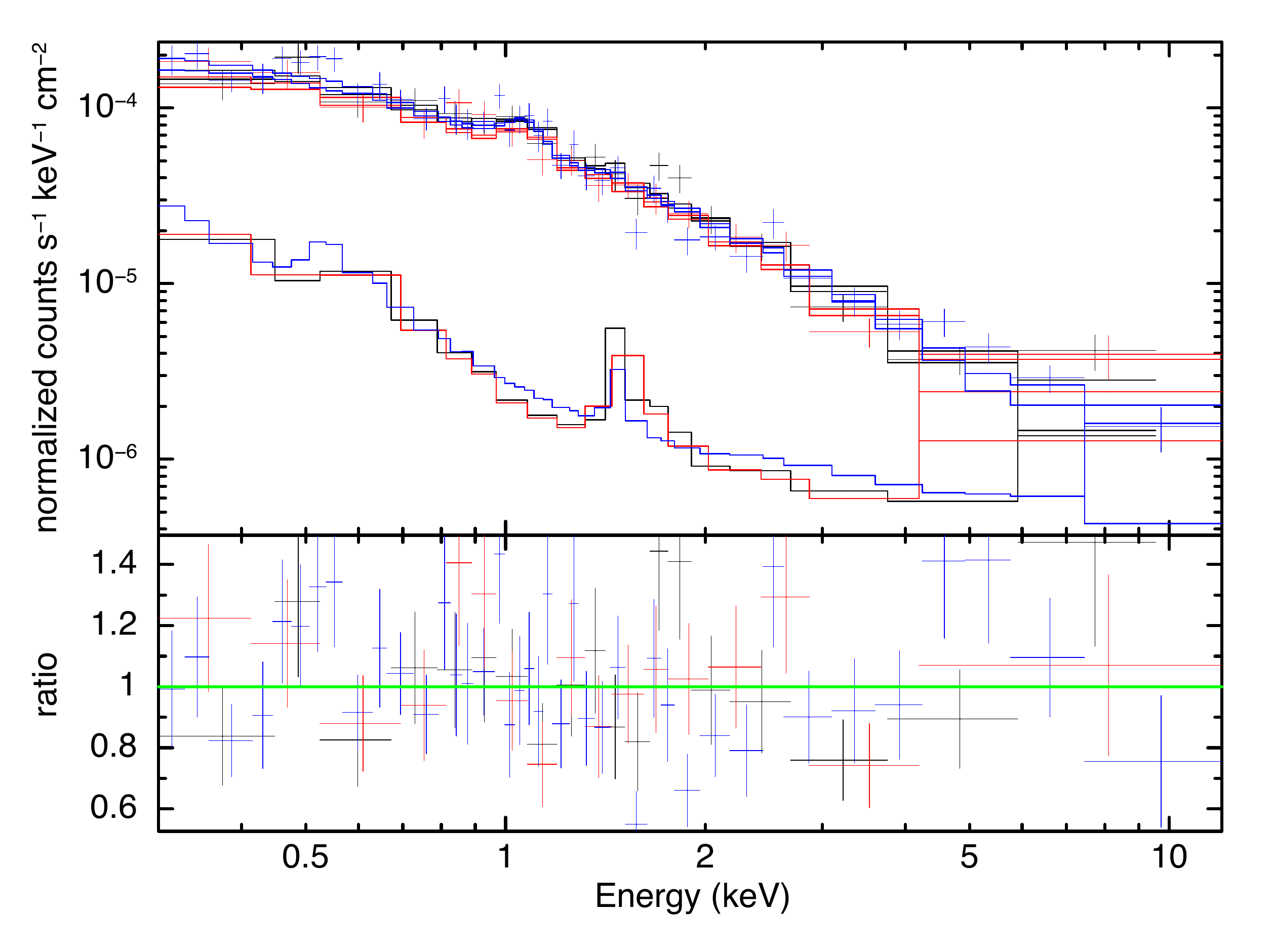}}
{\includegraphics[width=70mm]{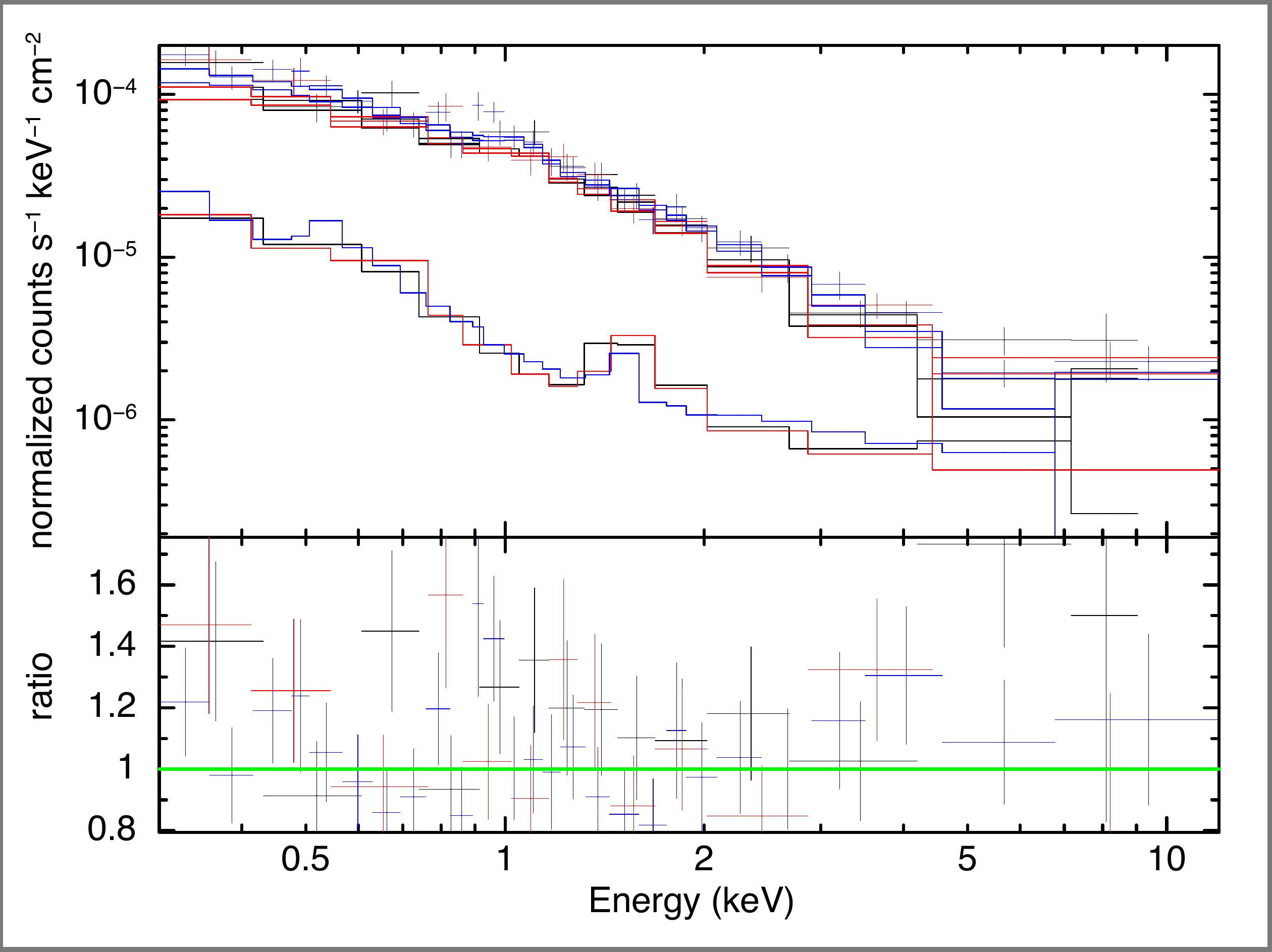}}
\caption{Same as Fig.~\ref{fig:Sectors}, but for \textit{XMM-Newton} EPIC spectra.}
\label{fig:SectorsXMM}
\end{figure}

\end{appendix}

\end{document}